\documentclass[aps,prb,floatfix,superscriptaddress,twocolumn]{revtex4-2}

\usepackage{graphicx}
\usepackage[english]{babel}
\usepackage{amsmath}
\usepackage{amssymb}
\usepackage{tensor}
\usepackage{xcolor}

\newcommand{\be}{\begin{eqnarray}}
\newcommand{\ee}{\end{eqnarray}}
\newcommand{\p}{\partial}

\newcommand{\dc}{c^{\dagger}}

\def\ket#1{|#1\rangle}

\def\ep#1{\langle #1 \rangle}

\begin{document}

\title{Structure and scaling of Kitaev chain across a quantum critical point in real space}

\author{Yan He}
\affiliation{College of Physics, Sichuan University, Chengdu, Sichuan 610064, China}
\email{heyan$_$ctp@scu.edu.cn}

\author{Chih-Chun Chien}
\affiliation{Department of Physics, University of California, Merced, CA 95343, USA.}
\email{cchien5@ucmerced.edu}

\begin{abstract}
The spatial Kibble-Zurek mechanism (KZM) is applied to the Kitaev chain with inhomogeneous pairing interactions that vanish in half of the lattice and result in a quantum critical point separating the superfluid and normal-gas phases in real space. The weakly-interacting BCS theory predicts scaling behavior of the penetration of the pair wavefunction into the normal-gas region different from conventional power-law results due to the non-analytic dependence of the BCS order parameter on the interaction. The Bogoliubov-de Gennes (BdG) equation produces numerical results confirming the scaling behavior and hints complications in the strong-interaction regime. The limiting case of the step-function quench reveals the dominance of the BCS coherence length in absence of additional length scale. Furthermore, the energy spectrum and wavefunctions from the BdG equation show abundant in-gap states from the normal-gas region in addition to the topological edge states. 
\end{abstract}

\maketitle

\section{Introduction}
A quantum phase transition (QPT) occurs when a change of the Hamiltonian leads to different phases of matter in the ground state, see Refs.~\cite{RevModPhys.69.315,Vojta_2003,sachdev_2011,Carr_book,Dutta_book} for a review. The transverse-field Ising model (TFIM) serves as an exactly solvable model for a clear demonstration of a QPT as the magnetic field crosses a critical value and changes the magnetic order. On the other hand, the BCS theory of fermionic superfluid provides a mean-field description of off-diagonal long-range order induced by attractive interactions~\cite{Leggett}. Shutting off the pairing interaction at zero temperature turns the superfluid into a normal Fermi gas through a QPT. Similar to conventional phase transitions, universal scaling behavior emerges near a QPT with diverging correlation length and time.

While conventional ways of studying QPTs focus on homogeneous systems in the thermodynamic limit, here we will take a different route and investigate a QPT in real space induced by an inhomogeneous form of the Hamiltonian, where a spatially changing parameter results in a symmetric phase and a symmetry-broken phase separated by a quantum critical point (QCP) in real space. Explicitly, we will follow the framework of the spatial Kibble-Zurek mechanism (KZM)~\cite{Zurekptis,Dziarmaga-doiqpt,Damski_2009,Dziarmaga-rev,Lacki17} to analyze the remnant of the order parameter from the symmetry-broken phase into the symmetric phase when the whole system is in equilibrium. 
Previous studies of the spatial KZM on the inhomogeneous TFIM~\cite{Zurekptis,Dziarmaga-doiqpt} and atomic spinor gases~\cite{Damski_2009} have revealed interesting scaling behavior with exponents reflecting the critical exponents of the corresponding homogeneous systems. On the other hand, Ref.~\cite{Parajuli23} shows that the non-analytic expression of the order parameter of the BCS theory of a Fermi superfluid with $s$-wave pairing leads to more complex scaling behavior when a QPT in real space is induced by inhomogeneous pairing interactions.

Here we apply the spatial KZM to a $p$-wave Fermi superfluid, which in 1D has the form of the Kitaev chain~\cite{Kitaev-chain}. The inhomogeneous pairing interaction drops to zero in real space to generate a QCP separating the superfluid and normal Fermi gas on the two sides. A linear ramp of of the interaction introduces a frozen length scale~\cite{Dziarmaga-rev} through the slope of the ramp that combines with the BCS theory  to produce interesting scaling behavior on the normal-gas side. 
Another type of quenches of the interaction profiles, called the step-function quench, has a sudden drop of the parameter across the critical point in real space and may be viewed as a limiting case of the spatial quench as the ramping rate goes to infinity. A previous study~\cite{Parajuli23} of the step-function quench of $s$-wave BCS superfluids in continuum reveals that the BCS coherence length becomes the only relevant length scale in the weak-interaction regime. Moreover, proximity effects of a semi-infinite Kitaev chain with odd-frequency pairing encountering a topological critical point have been studied in Ref.~\cite{PhysRevB.101.024509}.

We will set up the Bogoliubov-de Gennes (BdG) equation~\cite{degennes-sc,BdG-book} of the Kitaev chain of a $p$-wave Fermi superfluid with inhomogeneous interaction profiles to analyze the structure, scaling behavior, and spectrum as the system exhibits a QCP in real space due to the vanishing of the interactions in part of the lattice.
The Kitaev chain is a paradigm of topological superconductors~\cite{Kane_TIRev,Zhang_TIRev,ChiuRMP} and may host Majorana bound states~\cite{Kitaev-chain,PhysRevLett.100.096407}. The BdG formalism allows us to analyze the energy spectrum of the Kitaev chain across a QCP in real space. We will show that, in addition to the topological edge states near zero energy due to the hard-wall confinement, the normal-gas region introduces abundant in-gap states localized in the region with zero pairing interaction. The decay of Majorana fermions and Andreev bound states across nanowires interfacing a superconducting and a normal regions has been studied in Refs.~\cite{PhysRevB.86.085408,PhysRevB.87.024515,PhysRevB.98.245407}. Here we will focus on the decay of the pairing correlation.

Before presenting our studies, we remark on the difference between the time-independent KZM studied here and the time-dependent KZM, which was the original idea of analyzing structures across phase transitions~\cite{kibble-1976,kibble-1980,Zurek-1985} and is more common in the literature~\cite{Zurek-1996,Zurek-1997,Zurek-1999,Zurek-2002,Dziarmaga-rev}. For the time-dependent KZM, the system is homogeneous but driven out of equilibrium by a linear ramp of a parameter in time of the form $t/\tau_Q$. The scaling is determined by the quench rate $1/\tau_Q$, and topological defects may emerge in such nonequilibrium settings~\cite{QPT-Dziarmaga-2005,QPT-Zurek-2005,QIM-Jaschke-2017,Dziarmaga-2022}. In contrast, the spatial KZM keeps the system in equilibrium but introduces a linear ramp of a parameter in real space of the form $\alpha (x-x_c)$, where $x_c$ is the location of the transition and $\alpha$ is the quench rate in space. To establish equilibrium between the two sides of the critical point in real space produced by the spatially varying parameter, scaling behavior depending on $\alpha$ will emerge in the time-independent KZM and reveal properties of the corresponding homogeneous systems. More contrasts between time-dependent and time-independent KZMs can be found in Refs.~\cite{Dziarmaga-rev,Parajuli23}. While there have been many theoretical~\cite{PhysRevD.81.025017,Uhlmann_2010,adiabtic-dynamics-2005,QD-Legget-2005,PhysRevA.97.033626,PhysRevA.75.023603,PhysRevB.86.144521,PhysRevB.95.104306,PhysRevA.103.013310,PhysRevB.106.224302} and experimental~\cite{Monaco-2002,Ulm-2013,Pyka-2013,ALGaunt-2015,QPT-Brawn-2015,QPT-mott-2011,QKZM-Nature-2019,QKZM-prl-2016,PRXQuantum.4.010302,KZM-colloidal-monolayer-2015,KZM-FermiSF-2019,KZM-vortices-2021} studies on the time-dependent KZM, the time-independent KZM has been less explored and awaits more investigations. 

The rest of the paper is organized as follows. Sec.~\ref{Sec:Theory} briefly summarizes the Kitaev chain in the homogeneous case and the BdG formalism for solving an inhomogeneous Kitaev chain. Sec.~\ref{Sec:Quench} introduces the spatial and step-function quenches and reviews their scaling mechanisms. Sec.~\ref{Sec:Results} presents the profiles, scaling analyses, energy spectrum, and eigenfunctions of the Kitaev chain with inhomogeneous pairing interactions. The agreements with the weak-interaction predictions and deviations when the interaction is strong are demonstrated. We also discuss physical implications and relevance to previous works.
Finally, Sec.~\ref{Sec:Conclusion} concludes our work.

\section{Theoretical background of Kitaev chain}\label{Sec:Theory}
\subsection{Homogeneous systems}
The Kitaev chain is a model of a  spinless Fermi superfluid with $p$-wave paring. The Hamiltonian in real space is given by
\be\label{eq:H1}
H&=&\sum_i\Big[-w(\dc_i c_{i+1}+\dc_{i+1}c_i)-\mu\dc_ic_i+ \nonumber \\
& &\Delta(\dc_i\dc_{i+1}+c_{i+1}c_i)\Big].
\ee
Here $c_i$ and $c_i^{\dagger}$ are the fermion annihilation and creation operators on site $i$, $w$ is the hopping coefficient, and $\Delta$ is the gap function of the nearest-neighbor pairing. We set $\hbar=1=k_B$ throughout the paper and use $w$ and the lattice constant $a$ as the energy and length units, respectively. 

For a homogeneous system with periodic boundary condition, the Hamiltonian in momentum space in terms of the Nambu spinor $\psi=(c_k, \dc_{-k})^T$ is given by
\be
H&=&\left(\begin{array}{cc}
         -2w\cos k-\mu & -2i\Delta\sin k\ \\
         2i\Delta\sin k & 2w\cos k+\mu
\end{array}\right) \nonumber \\
&=&d_3\sigma_3+d_2\sigma_2.
\ee
Here $d_3=-2w\cos k-\mu$, $d_2=2\Delta\sin k$, and $\sigma_{i}$ with $i=1,2,3$ are the Pauli matrices applying to the Nambu space. The  Hamiltonian satisfy the particle-hole symmetry
\be
\sigma_1 H^*(-k) \sigma_1=-H(k).
\ee
Therefore, this model belongs to the class D, which has a $\mathbb{Z}_2$ index in 1D~\cite{ChiuRMP}.
The eigenvalues of the above Hamiltonian are given by
\be
E=\pm E_k,\quad E_k=\sqrt{d_2^2+d_3^2},\quad
\ee
The eigenstate of the lower band can be written as
\be
\ket{\psi}=\Big(v_k,\,i\,\textrm{sgn}(d_2)u_k\Big)^T.
\label{eq-uv}
\ee
Here $u_k=\sqrt{\frac{E_k+d_3}{2E_k}}$ and $v_k=\sqrt{\frac{E_k-d_3}{2E_k}}$.

The gap function is defined as $\Delta=g\ep{c_i c_{i+1}}$, where $g$ is the coupling constant of the attraction between the fermions. Following the BCS theory, the gap equation for a system of size $L$ at $T=0$ is given by
\be
\frac1g=\frac{1}{L}\sum_k\frac{2\sin^2 k}{E_k}.
\ee
For the half-filling case, $\mu=0$ and the right hand side becomes
\be
&&\frac{1}{2\pi}\int_0^{2\pi}\frac{\sin^2k dk}{\sqrt{w^2\cos^2 k+\Delta^2\sin^2 k}}\nonumber\\
&&=\frac{2}{\pi w(1-\Delta^2/w^2)}\Big[K\Big(\sqrt{1-\frac{\Delta^2}{w^2}}\Big)-E(\sqrt{1-\frac{\Delta^2}{w^2}})\Big].
\ee
Here we have used the complete elliptic integral
\be
K(q)&=&\int_0^{\pi/2}\frac{dx}{\sqrt{1-q^2\sin^2x}}, \nonumber \\
E(q)&=&\int_0^{\pi/2}\sqrt{1-q^2\sin^2x}dx.
\ee
For $q'=\sqrt{1-q^2}\ll1$, 
\be
K(q)=\ln\frac{4}{q'}+O((q')^2),\quad E(q)=1+O((q')^2).
\ee
Since $\Delta\ll w$ in the weakly-interacting regime, the gap equation becomes
\be
\frac1g
\approx\frac{2}{\pi w}\Big(\ln\frac{4w}{\Delta}-1\Big).
\ee
The gap function at $T=0$ in the weakly-interacting regime thus has the expression
\be\label{eq:Delta}
\Delta\approx 4w \exp(-\frac{\pi w}{2g}-1).
\ee
Since $\Delta$ is non-analytic in $g$, interesting scaling behavior will emerge in our study of inhomogeneous systems.

The BCS coherence length at $T=0$ is roughly the size of a Cooper pair~\cite{Leggett}, which is given by
\be
\xi_0^2=\frac{\ep{\psi_k|-\nabla_k^2|\psi_k}}{\ep{\psi_k|\psi_k}}.
\ee
Here $\psi_k=u_kv_k=\frac{\Delta\sin k}{2E_k}$ is the Cooper pair wavefunction, and $u_k$, $v_k$ are giving below Eq.~(\ref{eq-uv}). For the superfluid with $p$-wave pairing, we have
\be
&&\sum_k \ep{\psi_k|-\nabla_k^2|\psi_k}=\sum_k\Big(\frac{d\psi_k}{dk}\Big)^2=\frac{\pi(w^2+3\Delta^2)}{16w\Delta},\\
&&\sum_k \ep{\psi_k|\psi_k}=\sum_k\psi_k^2=\frac{\pi\Delta}{2(\Delta+w)}.
\ee
Thus,
\be\label{eq:BCSCohLength}
\xi_0\approx\frac{w}{2\sqrt{2}\Delta}.
\ee
Combining with Eq.~\eqref{eq:Delta}, one obtains the following scaling behavior of the BCS coherence length for fixed $w$:
\be\label{eq:StepQScaling}
\ln(\xi_0)\sim g^{-1}.
\ee

The topology of the Kitaev chain is characterized by the Berry phase in 1D, which is computed as
\be
\theta&=&i\int_0^{2\pi}dk \ep{\psi|\frac{\p}{\p k}|\psi}\nonumber \\
&=&\frac i2\int_0^{2\pi}dk \frac{\p}{\p k}\ln[\textrm{sgn}(d_2)]\Big(1+\frac{d_3}{E_k}\Big).
\ee
One can show that
\be\label{eq:TopoIndex}
\theta=\left\{
    \begin{array}{ll}
      \pi, & 2w>\mu,\\
      0, & 2w<\mu.
    \end{array}
  \right.
\ee
The bulk-boundary correspondence of topological systems~\cite{Kane_TIRev,Zhang_TIRev,ChiuRMP} relates the topological index in the bulk with the localized edge states at the boundary. 
Localized edge states with zero eigen-energy appear in the Kitaev chain with open boundary condition when the Berry phase of the corresponding system with periodic boundary condition takes the nontrivial value $\theta=\pi$. When $\theta=0$ in a periodic Kitaev chain, the system is topologically trivial and there is no localized edge state in the corresponding open chain.
We mention that the Kitaev chain of Eq.~(\ref{eq:H1}) has both particle-hole and time-reversal symmetries and therefore belongs to the BDI symmetry class~\cite{ChiuRMP}. Its topological index can be any integer. However, for the specific model of Eq.~(\ref{eq:H1}), the number of localized edge states can only be zero or one, which corresponds to the index of Eq.~(\ref{eq:TopoIndex}).
We mention that there are studies on thermal transport~\cite{HePLA23} and periodic driving~\cite{PhysRevB.98.125129} of the Kitaev chain out of equilibrium.

\begin{figure}
\centering
\includegraphics[width=\columnwidth]{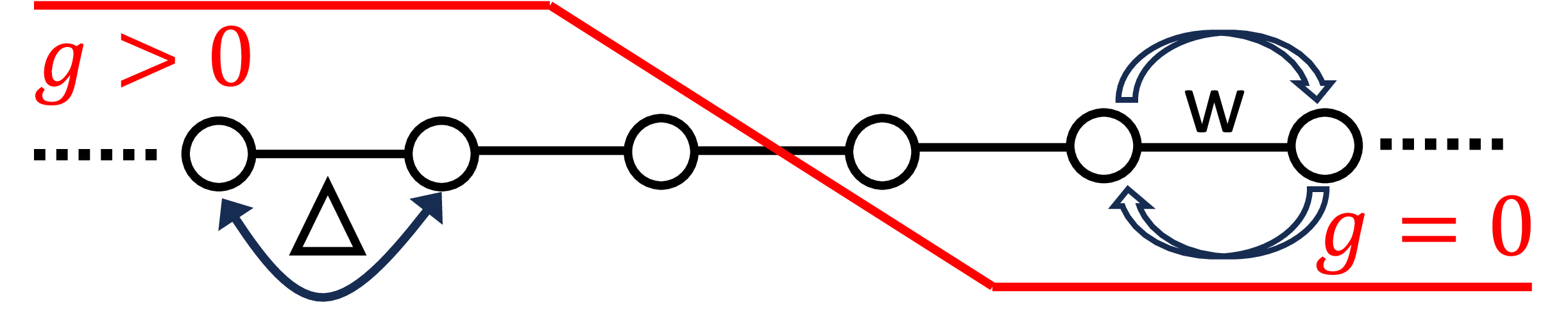}
\caption{Illustration of the Kitaev chain with inhomogeneous pairing interactions (red solid line). A linear ramp (sudden drop) corresponds to a spatial (step-function) quench.  The nearest-neighbor pairing gap $\Delta$ and hopping coefficient $w$ are also shown.}
\label{fig:illustration}
\end{figure}

\subsection{BdG equation for inhomogeneous cases}
Here we consider a Kitaev chain with a spatially dependent coupling constant $g_i$, as illustrated in Fig.~\ref{fig:illustration}. The Hamiltonian is similar to Eq.~\eqref{eq:H1} 
with a spatially dependent gap function 
\be\label{eq:Ddefinition}
\Delta_i=g_i\ep{c_i c_{i+1}}.
\ee
Explicitly,
\be \label{eq-TV}
&&H=\sum_{i,j}\Big[T_{ij}\dc_i c_j+\frac12V_{ij}\dc_i\dc_j+\frac12V_{ij}c_j c_i\Big],
\ee
where
$T_{ij}=-w(\delta_{i+1,j}+\delta_{i,j+1})-\mu\delta_{ij}$ and $V_{ij}=\Delta_i(\delta_{i+1,j}-\delta_{i,j+1})$.
The Hamiltonian can be diagonalized by a real-space Bogoliubov transformation \cite{Lieb61} via
\be
\eta_n&=&\sum_j \Big(u_{n,j}c_j+v_{n,j}\dc_j\Big), 
\eta^\dag_n=\sum_j \Big(u_{n,j}\dc_j+v_{n,j}c_j\Big). \nonumber \\
\label{eq-eta}
\ee
Here $u_n$ and $v_n$ are real coefficients. In order to have $\eta_n$ as fermion operators, the following orthonormal relations are imposed: 
\be
& &\sum_j\Big(u_{m,j}u_{n,j}+u_{m,j}u_{n,j}\Big)=\delta_{mn}, \nonumber \\
& &\sum_j\Big(u_{m,j}v_{n,j}+v_{m,j}u_{n,j}\Big)=0.
\ee

The Hamiltonian after the diagonalization becomes
\be
H=\sum_n E_n\eta^\dag_n\eta_n.
\label{eq-HE}
\ee
The anti-commutation relations between $\eta_n$ then lead to 
\be
[\eta_m,\,H]=E_m\eta_m.
\label{eq-etaH}
\ee
Substituting Eq.~(\ref{eq-TV}) and Eq.~(\ref{eq-eta}) into Eq.~(\ref{eq-etaH}), we arrived at the BdG equations 
\be\label{eq:BdG_L}
\sum_j\Big(T_{ij}u_{n,j}+V_{ij}v_{n,j}\Big)=E_n u_{n,j},\\
\sum_j\Big(V_{ji}u_{n,j}-T_{ij}v_{n,j}\Big)=E_n v_{n,j}.
\ee
The BdG equation has a particle-hole symmetry that pairs each $E_n > 0$ state with a $E_n < 0$ state, which allows us to further simplify the expressions.

Making use of the inverse Bogoliubov transformation
\be
c_j&=&\sum_n \Big(u_{n,j}\eta_n+v_{n,j}\eta^\dag_n\Big),~
c^\dag_j=\sum_n \Big(u_{n,j}\eta^\dag_n+v_{n,j}\eta_n\Big), \nonumber \\
\ee
we find that the pairing gap is determined by
$\Delta_i=\sum_n \Big[g_i u_{n,i}v_{n,i+1}[1-f(E_n)]+g_i u_{n,i+1}v_{n,i}f(E_n)\Big]$.
Here $f(x)=1/(e^{x/T}+1)$ is the Fermi function. At $T=0$, the gap function at site $i$ is given by
\be
\Delta_i=\sum_n \Big[g_i u_{n,i}v_{n,i+1}\Theta(E_n)+g_i u_{n,i+1}v_{n,i}\Theta(-E_n)\Big],\nonumber\\
\label{eq:BdG_Gap}
\ee
where $\Theta(x)$ is the step function. We remark that for the $p$-wave pairing of the Kitaev chain, the coefficients $u_n$, $v_n$ may be chosen real~\cite{Lieb61} to simplify the expressions.

The BdG equation is solved by iterations. A trial form of $\Delta_i$ and a given $\mu$ are plugged into Eq.~\eqref{eq:BdG_L}. After the diagonalization, the eigenvalues and eigenfunctions are found. A new gap function is then assembled according to Eq.~\eqref{eq:BdG_Gap}. A new iteration begins until the convergence condition $(1/L)\sum_i ||\Delta^{new}_i|-|\Delta^{old}_i|| < \epsilon$. Here $\Delta^{new}_i$ and $\Delta^{old}_i$ are the profiles of the gap function between two adjacent iterations and $L$ is the lattice size. We use $\epsilon=10^{-4}$ in our calculation. After the iteration converges, the ground-state density at site $i$ can be obtain by 
\be
n_i=\langle \dc_i c_i\rangle=\sum_n \Big[u_{n,i}^2\Theta(E_n)+v_{n,i}^2\Theta(-E_n)\Big].
\ee

\section{Quench of interaction in space}\label{Sec:Quench}
When the pairing coupling constant $g$ vanishes, the ground state of the Kitaev chain changes from a superfluid to a spin-polarized normal Fermi gas. With spatially inhomogeneous interactions, a quantum transition in real space may emerge as $g$ drops to zero. While many possible inhomogeneous profiles of the interactions may be postulated, here we follow Ref.~\cite{Parajuli23} and investigate two types of interaction quenches in real space illustrated in Fig.~\ref{fig:illustration}. The spatial quench has a linear ramp of the interaction while the step-function quench has an abrupt drop. The formal may apply to systems with a gradual change of the pairing interactions as the system transits from the superfluid to the normal gas while the latter models systems with a distinct interface between the two regions.

Since the gap function shown in Eq.~\eqref{eq:BdG_Gap} vanishes when $g_i=0$, it does not reflect the penetration of the pairing into the normal-gas phase. Instead, we analyze the pair wavefunction $F_j=\ep{c_jc_{j+1}}$, which decays into the normal-gas phase and exhibits interesting scaling behavior.

\subsection{Spatial quench}
We consider a linear ramp of the pairing interaction, which allows for an analysis of the scaling behavior according to the spatial KZM. The coupling constant $g$ linearly drops to zero inside the interval $\frac{L}2-L_1<i<\frac{L}2$ with the profile 
\be
g_i=\left\{
      \begin{array}{ll}
        g_0, & 1\le i\le L/2-L_1, \\
        g_0\dfrac{L/2-i}{L_1}, & L/2-L_1\le i\le L/2, \\
        0, &   L/2<i\le L.
      \end{array}
    \right.
\label{eq-gi-1}
\ee
As a consequence, a real-space QCP at $x_c=L/2$ emerges as $g_i$ drops to zero, separating the superfluid phase on the left and the normal phase on the right. The ramp of the interaction profile introduces a slope $\alpha = g_0/(L_1)$, which plays the role of the quench rate of the time-dependent KZM~\cite{Dziarmaga-rev}.

Near the QCP at $x_c=L/2$, we may write $g_0-\alpha(x-x_c) = -\alpha y$, where $y=(x-x_c-g_0/\alpha)$. In the ramp, the coherence length freezes at $\xi=\xi_{fr}$ due to the drop of the interaction, which then introduces the frozen interaction strength $g_{fr}\sim \alpha\xi_{fr}$. From Eq.~\eqref{eq:Delta} in the weakly interacting regime, the frozen coherence length leads to $\Delta_{fr}\sim\exp(-1/g_{fr})\sim\exp(1/(\alpha \xi_{fr}))$. Since the ramp occurs in the superfluid regime, the BCS coherence length given by Eq.~\eqref{eq:BCSCohLength} leads to a consistent equation for $\xi_{fr}$, given by $\xi_{fr}\sim (1/\Delta_{fr})\sim\exp(-1/(\alpha \xi_{fr}))$. The frozen coherence length determines the correlation on the other side of the QCP, so the characteristic length $\xi_F$ of the pair wavefunction on the normal-gas region is determined by $\xi_{fr}$. Therefore, we obtain the scaling
\be\label{eq:xiFsq}
\xi_{F}\ln(\xi_F)\sim 1/\alpha.
\ee

The result is different from the power-law scaling behavior of the spatial KZM in magnetic systems~\cite{Zurekptis,Dziarmaga-doiqpt} because the order parameter of the BCS theory, which is the gap function, is nonanalytic in the interaction strength even in the weakly interacting regime. The correlation length is thus dominated by the nonanalytic behavior and exhibits interesting scaling behavior as the system crosses a superfluid to normal-gas QCP in real space. We mention if the correlation length diverges near the critical point according to a power law $\xi\sim \epsilon^{-\nu}$ like the TFIM, where $\epsilon$ measures the distance to the transition, the freezing length $\xi_{fr}$ occurs when $\epsilon_{fr}\sim\alpha\xi_{fr}$, where $\alpha$ is the slope of the spatial quench. Therefore, $\xi\sim \alpha^{-\nu/(1+\nu)}$ for the conventional  characteristic length from the spatial KZM. We remark that if the non power-law scaling is viewed as a limiting case with $\nu\rightarrow\infty$, as proposed in Ref.~\cite{PhysRevB.95.104306} for some particular types of time-dependent KZM, the argument will lead to $\xi\sim 1/\alpha$, which differs from Eq.~\eqref{eq:xiFsq} by a logarithmic correction.

\subsection{Step-function quench}
The step-function quench with a sudden drop of the coupling constant $g$ in the middle of the chain may be considered as a limiting case of the spatial quench when the ramp is infinitely narrow. Explicitly, the location-dependent coupling constant  follows the equation 
\be
g_i=\left\{
      \begin{array}{ll}
        g_0, & 1\le i\le L/2, \\
        0, &   L/2<i\le L.
      \end{array}
    \right. 
\label{eq-gi}
\ee
When $g_i$ drops to zero, the system transitions from a superfluid to a normal gas. Therefore, a QCP in real space is present at $x_c=L/2$.
The sudden drop of the pairing interaction leaves no additional length scale other than the BCS coherence length for the fermion pairs. Therefore, the characteristic length $\xi_F$ of the penetration of the pair wavefunction $F$ in the normal region is expected to follow the scaling relation 
\be\label{eq:xiF_SFQ}
\ln(\xi_F)\sim g_0^{-1}
\ee
in the weak interaction regime similar to that of the BCS coherence length of Eq.~\eqref{eq:BCSCohLength} in the bulk of the superfluid region.

\begin{figure}
\centering
\includegraphics[width=0.7\columnwidth]{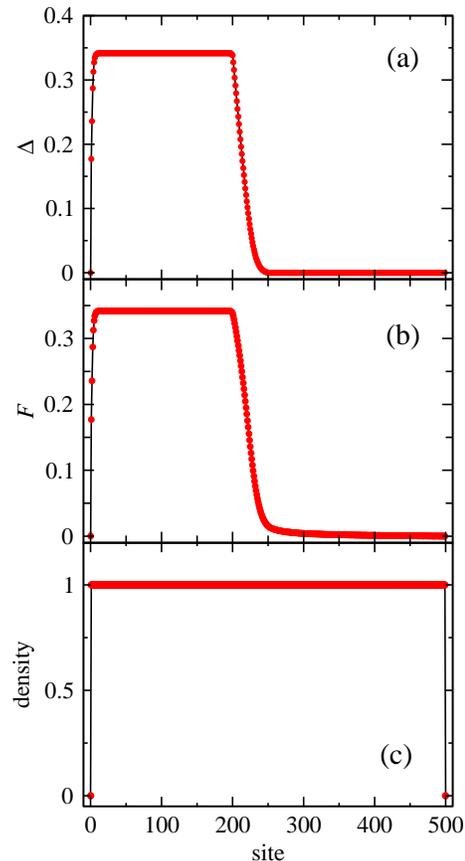}
\caption{Profiles of (a) the gap function, (b) pair wavefunction $F_i=\ep{c_ic_{i+1}}$, and (c) density as functions of the site index of a spatial quench.
Here $g_0=1$, $\mu=0$, $L=500$, and the ramp width is $L_1=50$. }
\label{fig-g2}
\end{figure}

\begin{figure}
\centering
\includegraphics[width=0.8\columnwidth]{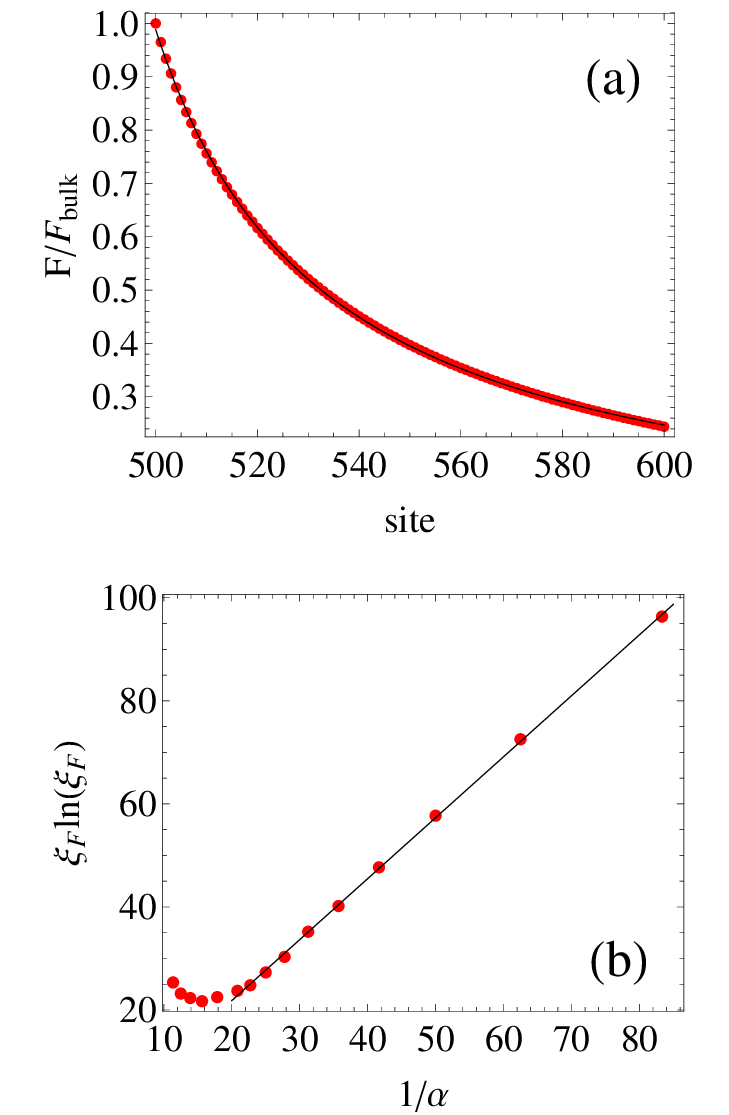}
\caption{(a) Fitting of the pair wavefunction $F$ by Eq.~\eqref{eq:F_fit} in the normal-gas region in a spatial quench. Here $g_0=1$, $\mu=0$, $L=1000$, and the ramp width is $L_1=100$. (b) The characteristic length $\xi_{F}$ extracted from the fitting of $F$ in the normal-gas region as a function of $1/\alpha$ (red dots). The black line is Eq.~\eqref{eq:xiFsq}. A deviation from the weak-interaction prediction is visible when $1/\alpha$ decreases.
}
\label{quench-fit}
\end{figure}

\section{Results and Discussions}\label{Sec:Results}
\subsection{Scaling analyses}
Here we show our solutions of the BdG equation with open boundary condition. Up to $1000$ sites in the system have been analyzed to obtain the scaling behavior of the pair wavefunction in the normal-gas region. 
We first show the profiles of the gap function, $F_j$, and density as functions of the site index for a selected case undergoing a spatial quench in Fig. \ref{fig-g2}. 
Due to the vanishing $g_i$ at $x_c=L/2$, the gap function drops to zero at $x_c$ as well. However, the pair wavefunction continuously extends into the normal-gas region. On the other hand, the density is basically flat across the QCP in real space, as pairing in the BCS regime does not introduce drastic changes to the expression of the density.
The introduction of a linear ramp in the interaction profile results in interesting scaling behavior, which we will analyze here.

To fit the decay of the pair wavefunction $F$ in the normal-gas region, we refine the grid and zoom in the right-half of the solutions.  For $s$-wave pairing in continuum, it has been shown~\cite{degennes-proximity,Falk-PE,PE-Silvert} that the decay of $F$ in the normal-gas region at zero temperature follows the power-law form
\be\label{eq:F_fit}
F(x)\sim\frac{\xi_F}{x-x_c},
\ee 
which also defines the characteristic length $\xi_F$ in an inhomogeneous Fermi superfluid. In other words, $\xi_F$ measures the typical range of penetration of the pair wavefunction $F$.
Although here we study $p$-wave pairing in a lattice system, we found the power-law form still fits our data well, as shown in Fig.~\ref{quench-fit} (a). The fitting allows us to extract the characteristic length $\xi_F$ for a particular set of parameters. After collecting more values of $\xi_F$ as a function of the ramp slope $\alpha=g_0/L_1$  of the pairing interactions, we compare the data with the functional form of Eq.~\eqref{eq:xiFsq} from the spatial KZM in the weak-interaction regime, as shown in Fig.~\ref{quench-fit} (b). The agreement between the numerical simulations and analytic formula  in the weak-interaction regime shows the universal behavior of the spatial KZM applied to $p$-wave Fermi superfluids described by the mean-field BCS theory in the ground state. Away from the weak-interaction regime when $1/\alpha$ becomes small, however, the characteristic length $\xi_F$ starts to deviate from the weak-interaction prediction because the analytic expression~\eqref{eq:xiFsq} no longer applies although the fitting of $F$ by Eq~\eqref{eq:F_fit} still works.

\begin{figure}
\centering
\includegraphics[width=0.7\columnwidth]{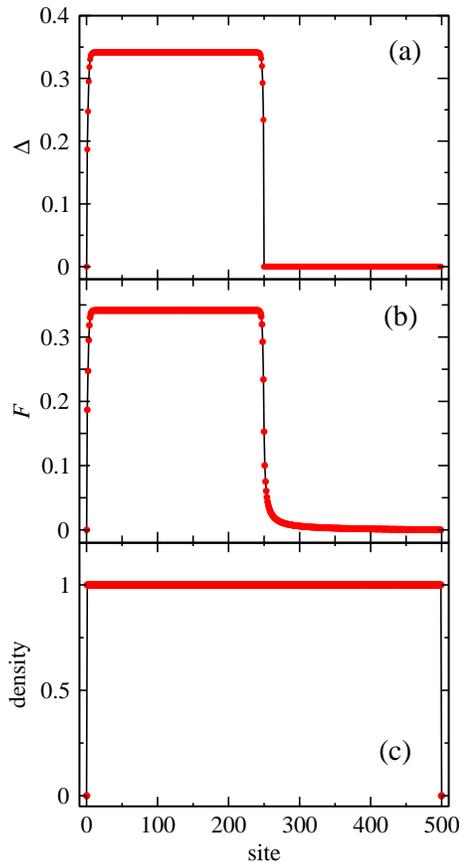}
\caption{Profiles of (a) the gap function, (b) pair wavefunction $F_i=\ep{c_ic_{i+1}}$, and (c)  density as functions of the site index for a step-function quench of Eq.~(\ref{eq-gi}). Here we assume $g_0=1$, $\mu=0$, and the system size is $L=500$.}
\label{fig-g1}
\end{figure}

On the other hand, the profiles of the gap function $\Delta$, pair wavefunction $F_j=\ep{c_jc_{j+1}}$, and density as functions of the site index following a step-function quench are show in Fig. \ref{fig-g1}. 
At first look, they are very similar to their counterparts in the spatial quench. However, the vanishing of the width of the ramp leads to a different scaling analysis for the step-function quench because there is no longer an additional length scale (from the slope) as the system crosses the QCP in real space.

\begin{figure}
\centering
\includegraphics[width=0.8\columnwidth]{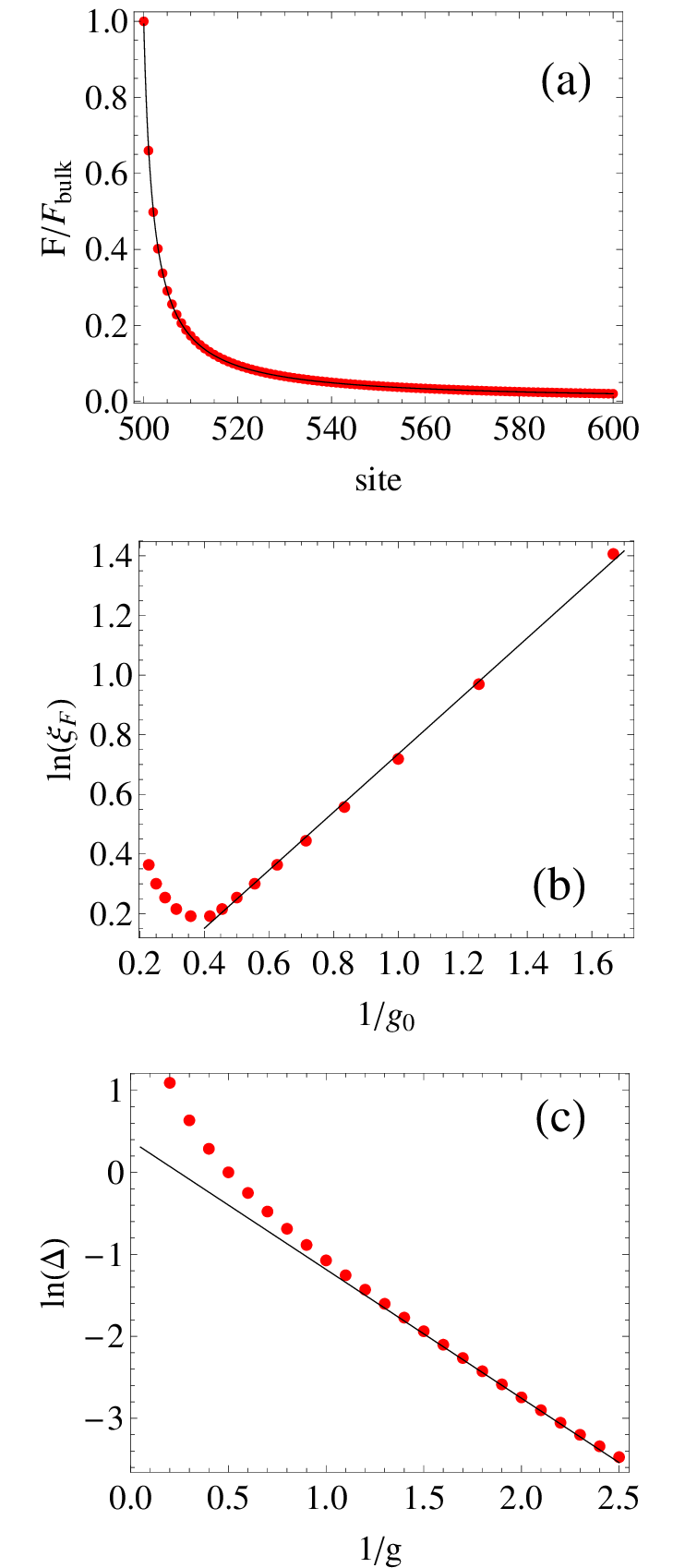}
\caption{(a) Fitting of the  pair wavefunction $F$ in the normal-gas region by Eq.~\eqref{eq:F_fit}. Here $g_0=1$, $\mu=0$ with $L=1000$. (b)  $\ln\xi_F$  as a function of $1/g_0$ (red dots) and the BCS coherence length of the superfluid region in the weak-interaction regime (black line). (c) $\ln(\Delta)$ as a function of $1/g$ for a uniform Kitaev chain with $\mu=0$ (red dots). The black line shows Eq.~\eqref{eq:Delta} from the weak-interaction regime. Deviations from the weak-interaction predictions are observable in (b) and (c).
}
\label{fig-fit}
\end{figure}

The fitting of $F$ in the normal-gas region of a selected step-function quench case with the form of Eq.~\eqref{eq:F_fit} is shown in  Fig. \ref{fig-fit} (a), which allows us to extract the value of the characteristic length $\xi_F$ for this set of parameters.
By extracting $\xi_F$ as a function of $g_0$, we show the scaling behavior of $\xi_F$ in Fig~\ref{fig-fit} (b). One can see that the scaling agrees with Eq.~\eqref{eq:xiF_SFQ} in the weak-interaction regime. Therefore, the BCS coherence length on the superfluid side determines the penetration of the pair wavefunction into the normal-gas region because there is no additional length scale in a step-function quench across the QCP in real space. As $g_0$ increases, $\xi_F$ shows a deviation from the weak-interaction prediction. The reason is because the BCS gap function deviates from the weak-interaction expression~\eqref{eq:Delta} as $g_0$ increases since we have verified the fitting of $F$ by Eq.~\eqref{eq:F_fit} still works well when $g_0$ is large. To visualize the deviation of $\Delta$ from the weak-interaction expression, we show in Fig~\ref{fig-fit} (b) the gap function of a homogeneous Kitaev chain compared to the weak-interaction formula~\eqref{eq:Delta}.

\begin{figure}[t]
\centering
\includegraphics[width=0.7\columnwidth]{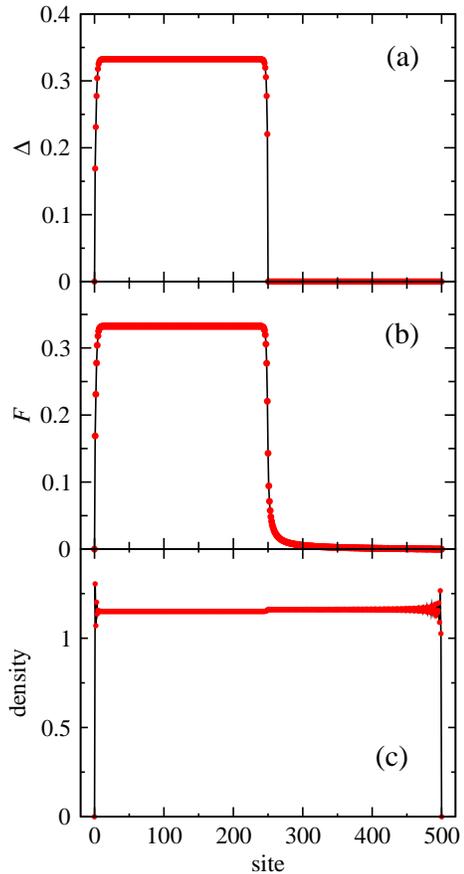}
\caption{Profiles of (a) $\Delta$, (b) $F$, and (c) density for a step-function quench with $\mu=0.5$. Here $g_0=1$ and $L=500$.}
\label{mu}
\end{figure}

We have two remarks about the results: (I) Our solutions to the  BdG equation only cover a limited range of the interaction or its slope. If $g_0$ or $\alpha$ is too small, the bulk gap becomes exponentially small due to Eq.~\eqref{eq:Delta}. It is difficult for the numerical iterations to converge. The small value of the pair wavefunction also makes it challenging to extract the correct value on the normal-gas side. On the other hand, if $g_0$ is too large or the slope is too steep, the convergence of the BdG equation becomes slow because a small adjustment may lead to a substantial change in the related quantities. Moreover, a strong pairing interaction on the left side pushes the system away from the BCS regime, making it challenging to have a continuous connection with the normal-gas region on the right. Nevertheless, we managed to obtain enough results to demonstrate the agreements of the scaling behavior  in the weak-interaction regime and the deviations from the weak-interaction predictions away from that regime.
(II) We have checked other values of $\mu$ and found that while there are quantitative changes in the density profile, the scaling behavior of both spatial quench and step-function quench remain the same as the $\mu=0$ case for a reasonable range of the interaction profiles. An illustration of the profiles of $\Delta$, $F$, and density for the case with $\mu=0.5$ in a step-function quench is shown in Fig. \ref{mu}. While there are wiggles in the density profile, the pair wavefunction remains smooth and its penetration in the normal-gas region follows the same scaling as that of the $\mu=0$ case, so we will not repeat the results here. Meanwhile, the larger value of $\mu$ leads to a slight mismatch of the densities between the superfluid and normal regions, as shown in Fig. \ref{mu} (c).

\subsection{Energy spectrum and eigenstates}
\begin{figure}[t]
\centering
\includegraphics[width=\columnwidth]{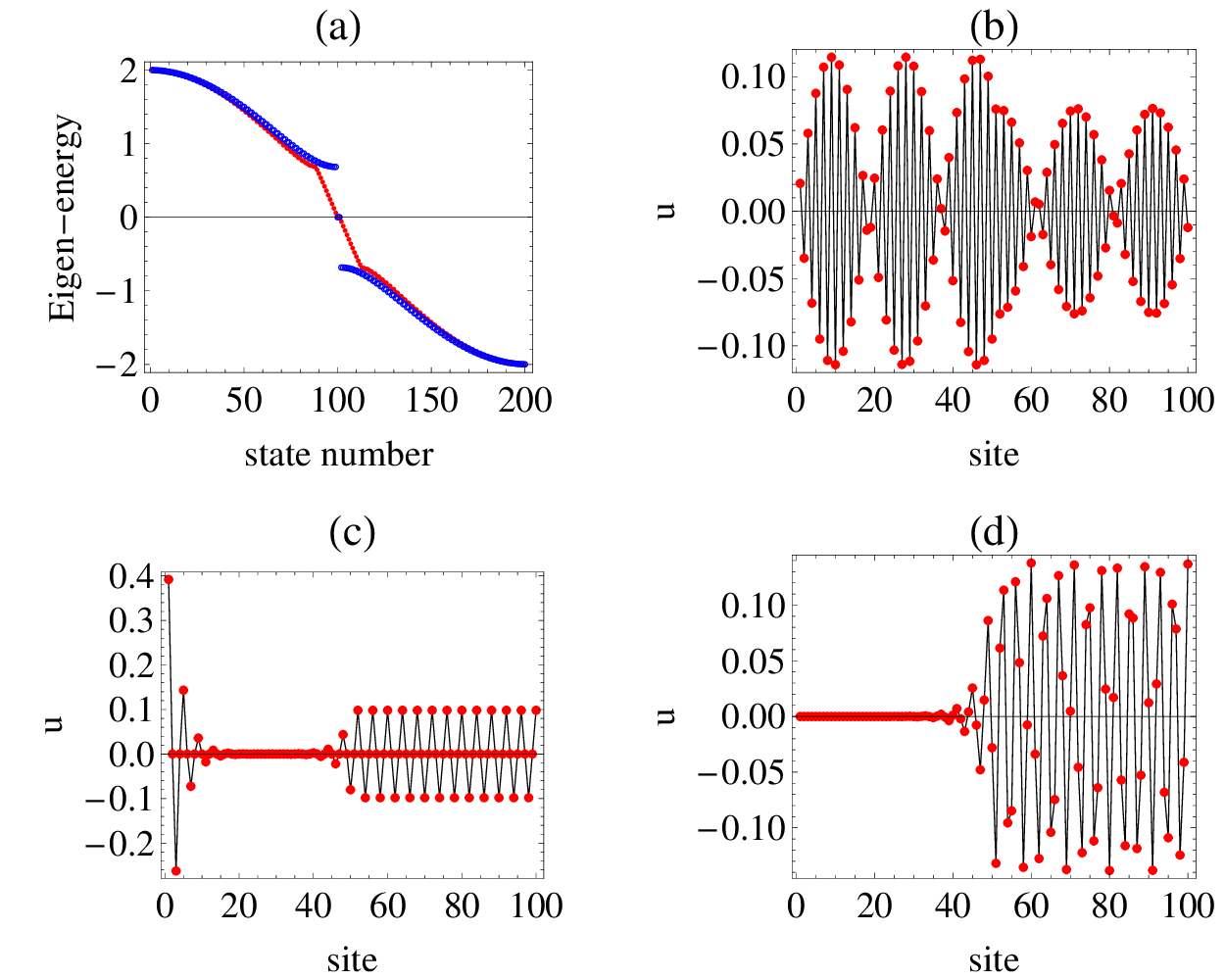}
\caption{(a) Energy spectra of a step-function quenched system (small red dots) and a homogeneous system (large blue circles). For the step-function quench, the wavefunction $u$ of a typical bulk state in the upper band, an edge state near zero energy, and a normal-gas state with $E\approx 0.3$ are shown in (b), (c), and (d), respectively.  Here $v$ behaves similarly as $u$, and $L=100$ for better visualization of the wavefunctions. 
}
\label{fig-ua}
\end{figure}

Here we investigate the  states of the Kitaev chain in the presence of a hard-wall box potential and spatially varying pairing interactions. 
The BdG equation allows us to analyze both the energy spectrum and the profiles of individual states. Panel (a) of Fig.~\ref{fig-ua} shows the eigen-energies and panels (b), (c), (d) plot some selected states for a step-function quench. 
To contrast the features from the inhomogeneous interaction profile with the homogeneous one, we plot in the same panel the energy spectrum of a similar system with a homogeneous interaction profile of $g_0$.
While the bulk bands are alike, one can see that there are many in-gap states for the step-function quench case. We will show that there are two types of in-gap states, one from the localization at the hard wall on the superfluid side and the other from the unpaired fermions on the right half of the chain.

For the step-function case, some selected wave functions $u_{n}$ in the bulk band, near zero energy, and inside the energy gap as functions of the site index from the BdG equation are shown in Fig.~\ref{fig-ua} (b), (c), (d). Panel (b) is a bulk state with its amplitude spreading over the whole chain. Panel (c) shows the state near $E=0$, which localizes at the left hard wall. The hard wall may be viewed as a boundary between the fermions and vacuum, so a localized edge state can be trapped there in the topological regime. This type of in-gap states is also present when the profile of $g_0$ is uniform. According to Eq.~\eqref{eq:TopoIndex}, the system is indeed topological when $\mu=0$ for the half-filling case. Finally, panel (d) shows a state inside the energy gap but away from $E=0$. It displays non-zero amplitude only in the region where $g_i=0$, thereby representing a normal-gas state due to the vanishing pairing interaction on the right half. We remark that the states shown in panels (c) and (d) have similar profiles except for the localized state on the left end in the zero-energy case.

\subsection{Implications}
While the spatial KZM of TFIM has been shown to exhibit power-law scaling of the decaying order parameter in the symmetric phase~\cite{Zurekptis,Dziarmaga-doiqpt}, the non-analytic behavior of the BCS theory leads to complicated scaling behavior, as pointed out in Ref.~\cite{Parajuli23} and here. Our results thus demonstrate the variety and versatility of the spatial KZM as a general framework. While the freezing of the coherence length due to the linear ramp applies to general types of phase transitions, the scaling behavior depends on the functional forms from the underlying systems. The peculiar scaling of the BCS theory across a QPT in real space thus provides another exotic feature of fermionic superfluids. We remark that if a bosonic system is considered instead, turning off the interaction between bosons leaves a Bose-Einstein condensate of noninteracting bosons at $T=0$, which is still considered as a symmetry-broken phase and makes it different from the fermionic case studied here.

There are several differences between the $p$-wave superfluid and the $s$-wave one studied in Ref.~\cite{Parajuli23}: (1) The $p$-wave pairing is between single-component (spin-polarized) fermions while the $s$-wave pairing is between two-component fermions. When physical systems are concerned, the $s$-wave case corresponds to a conventional superconductor next to a normal metal while the $p$-wave case corresponds to a triplet superconductor next to a ferromagnet. (2) The $p$-wave pairing is between adjacent lattice sites while the $s$-wave pairing is onsite. As a consequence, the pair wavefunction $F$ is onsite for $s$-wave pairing but between adjacent sites for $p$-wave pairing. (3) The Kitaev chain describes a topological superfluid with properties absent in the $s$-wave model. The topology of the Kitaev is expected to host Majorana bound states at the edge, as shown in Fig.~\ref{fig-ua} (c).

Nevertheless, we have shown that the functional forms of the order parameters represented by the gap functions and the BCS coherence lengths are similar for the $s$-wave and $p$-wave pairing cases in the weak-interaction regime, as a consequence of the BCS mean-field theory in the weak-pairing limit. Thus, their scaling behaviors are shown to be similar at least in that regime. The scaling analyses also show the universal applications of the frameworks of the spatial KZM and spatial quench to spatially inhomogeneous systems.

Meanwhile, it is easier to push the $p$-wave case away from the weak-pairing limit than its $s$-wave counterpart since the nearest-neighbor pairing of the $p$-wave case complicates the energy dispersion and gap function. This allows us to demonstrate the deviations of the Kitaev chain from the analytic expressions derived in the weak-pairing limit, as shown in Figs.~\ref{quench-fit} and \ref{fig-fit}. In contrast, Ref.~\cite{Parajuli23} did not explore the deviations from the weak-pairing regime due to numerical complications as it will take much stronger interactions to explore the deviations. Away from the weak-pairing regime, the $s$-wave and $p$-wave cases may exhibit more quantitatively different behaviors due to the different gap equations. We remark that the mean-field treatment of the Kitaev chain in this work renders it integrable, thereby escaping the eigenvalue thermalization hypothesis~\cite{Deutsch_2018}. Future studies of more complex systems with beyond mean-field interactions and disorder may connect to interesting phenomena such as many-body localization~\cite{Altman2018,RevModPhys.91.021001}.

There have been many attempts to realize the Kitaev chain, including those using quantum dots~ \cite{Dvir23} and theoretical work cautioning some subtleties~\cite{PhysRevB.107.035440} (see also Ref.~\cite{Oppen17} for a review). Moreover, quantum simulators or computers also offer insights into the Kitaev chain~\cite{Stenger21,Huang21,Mi22,Rancic2022,Iizuka2023}. Since the attractive interactions are induced or effective in most settings, it may be possible to tune the attraction in real space by varying the chemical or physical properties, such as the composition, electric or magnetic field, strain, light-matter interactions or other means, of the systems or simulators of the Kitaev chain to generate the inhomogeneous interactions studied in this work. Furthermore, $p$-wave pairing may also arise in possible triplet superconductivity induced by proximity effect in superconductor-ferromagnet heterostructures~\cite{SC-FM-rev}. Therefore, the penetration of superconducting correlation across a critical point in real space discussed here may apply to those composite systems if the interactions can be controlled by chemical compositions, lattice structures, or other means.

\section{Conclusion}\label{Sec:Conclusion}
The BdG equation of a $p$-wave Fermi superfluid described by the Kitaev chain with inhomogeneous pairing interaction profiles has revealed how the remnant of the pair wavefunction survives in the normal-gas region as the system in equilibrium exhibits a quantum critical point in real space. The mean-field BCS analysis leads to scaling behavior of the spatial and step-function quenches confirmed by the BdG equation in the weak- and intermediate- interaction regimes. The linear ramp of the spatial quench introduces an additional length scale from the slope, resulting in the spatial KZM, while the step-function quench is dominated by the BCS coherence length on the superfluid side. The energy spectrum and eigenfunctions further distinguish the topological edge states from the normal-gas states. Our study thus offers another example of using inhomogeneity to explore complex quantum systems.

\begin{acknowledgments} 
Y. H. was supported by the NNSF of China (No. 11874272) and Science Specialty Program of Sichuan University (No. 2020SCUNL210). C. C. C. was partly supported by the NSF (No. PHY-2310656).
\end{acknowledgments}


\begin{thebibliography}{74}%
	\makeatletter
	\providecommand \@ifxundefined [1]{%
		\@ifx{#1\undefined}
	}%
	\providecommand \@ifnum [1]{%
		\ifnum #1\expandafter \@firstoftwo
		\else \expandafter \@secondoftwo
		\fi
	}%
	\providecommand \@ifx [1]{%
		\ifx #1\expandafter \@firstoftwo
		\else \expandafter \@secondoftwo
		\fi
	}%
	\providecommand \natexlab [1]{#1}%
	\providecommand \enquote  [1]{``#1''}%
	\providecommand \bibnamefont  [1]{#1}%
	\providecommand \bibfnamefont [1]{#1}%
	\providecommand \citenamefont [1]{#1}%
	\providecommand \href@noop [0]{\@secondoftwo}%
	\providecommand \href [0]{\begingroup \@sanitize@url \@href}%
	\providecommand \@href[1]{\@@startlink{#1}\@@href}%
	\providecommand \@@href[1]{\endgroup#1\@@endlink}%
	\providecommand \@sanitize@url [0]{\catcode `\\12\catcode `\$12\catcode
		`\&12\catcode `\#12\catcode `\^12\catcode `\_12\catcode `\%12\relax}%
	\providecommand \@@startlink[1]{}%
	\providecommand \@@endlink[0]{}%
	\providecommand \url  [0]{\begingroup\@sanitize@url \@url }%
	\providecommand \@url [1]{\endgroup\@href {#1}{\urlprefix }}%
	\providecommand \urlprefix  [0]{URL }%
	\providecommand \Eprint [0]{\href }%
	\providecommand \doibase [0]{https://doi.org/}%
	\providecommand \selectlanguage [0]{\@gobble}%
	\providecommand \bibinfo  [0]{\@secondoftwo}%
	\providecommand \bibfield  [0]{\@secondoftwo}%
	\providecommand \translation [1]{[#1]}%
	\providecommand \BibitemOpen [0]{}%
	\providecommand \bibitemStop [0]{}%
	\providecommand \bibitemNoStop [0]{.\EOS\space}%
	\providecommand \EOS [0]{\spacefactor3000\relax}%
	\providecommand \BibitemShut  [1]{\csname bibitem#1\endcsname}%
	\let\auto@bib@innerbib\@empty
	\bibitem [{\citenamefont {Sondhi}\ \emph {et~al.}(1997)\citenamefont {Sondhi},
		\citenamefont {Girvin}, \citenamefont {Carini},\ and\ \citenamefont
		{Shahar}}]{RevModPhys.69.315}%
	\BibitemOpen
	\bibfield  {author} {\bibinfo {author} {\bibfnamefont {S.~L.}\ \bibnamefont
			{Sondhi}}, \bibinfo {author} {\bibfnamefont {S.~M.}\ \bibnamefont {Girvin}},
		\bibinfo {author} {\bibfnamefont {J.~P.}\ \bibnamefont {Carini}},\ and\
		\bibinfo {author} {\bibfnamefont {D.}~\bibnamefont {Shahar}},\ }\bibfield
	{title} {\bibinfo {title} {Continuous quantum phase transitions},\ }\href
	{https://doi.org/10.1103/RevModPhys.69.315} {\bibfield  {journal} {\bibinfo
			{journal} {Rev. Mod. Phys.}\ }\textbf {\bibinfo {volume} {69}},\ \bibinfo
		{pages} {315} (\bibinfo {year} {1997})}\BibitemShut {NoStop}%
	\bibitem [{\citenamefont {Vojta}(2003)}]{Vojta_2003}%
	\BibitemOpen
	\bibfield  {author} {\bibinfo {author} {\bibfnamefont {M.}~\bibnamefont
			{Vojta}},\ }\bibfield  {title} {\bibinfo {title} {Quantum phase
			transitions},\ }\href {https://doi.org/10.1088/0034-4885/66/12/R01}
	{\bibfield  {journal} {\bibinfo  {journal} {Rep. Prog. Phys.}\ }\textbf
		{\bibinfo {volume} {66}},\ \bibinfo {pages} {2069} (\bibinfo {year}
		{2003})}\BibitemShut {NoStop}%
	\bibitem [{\citenamefont {Sachdev}(2011)}]{sachdev_2011}%
	\BibitemOpen
	\bibfield  {author} {\bibinfo {author} {\bibfnamefont {S.}~\bibnamefont
			{Sachdev}},\ }\href {https://doi.org/10.1017/CBO9780511973765} {\emph
		{\bibinfo {title} {Quantum Phase Transitions}}},\ \bibinfo {edition} {2nd}\
	ed.\ (\bibinfo  {publisher} {Cambridge University Press},\ \bibinfo {address}
	{Cambridge, UK},\ \bibinfo {year} {2011})\BibitemShut {NoStop}%
	\bibitem [{\citenamefont {Carr}(2010)}]{Carr_book}%
	\BibitemOpen
	\bibinfo {editor} {\bibfnamefont {L.~D.}\ \bibnamefont {Carr}},\ ed.,\
	\href@noop {} {\emph {\bibinfo {title} {Understanding Quantum Phase
				Transitions}}}\ (\bibinfo  {publisher} {CRC Press},\ \bibinfo {address} {Boca
		Raton},\ \bibinfo {year} {2010})\BibitemShut {NoStop}%
	\bibitem [{\citenamefont {Dutta}\ \emph {et~al.}(2015)\citenamefont {Dutta},
		\citenamefont {Aeppli}, \citenamefont {Chakrabarti}, \citenamefont
		{Divakaran}, \citenamefont {Rosenbaum},\ and\ \citenamefont
		{Sen}}]{Dutta_book}%
	\BibitemOpen
	\bibfield  {author} {\bibinfo {author} {\bibfnamefont {A.}~\bibnamefont
			{Dutta}}, \bibinfo {author} {\bibfnamefont {G.}~\bibnamefont {Aeppli}},
		\bibinfo {author} {\bibfnamefont {B.~K.}\ \bibnamefont {Chakrabarti}},
		\bibinfo {author} {\bibfnamefont {U.}~\bibnamefont {Divakaran}}, \bibinfo
		{author} {\bibfnamefont {T.~F.}\ \bibnamefont {Rosenbaum}},\ and\ \bibinfo
		{author} {\bibfnamefont {D.}~\bibnamefont {Sen}},\ }\href@noop {} {\emph
		{\bibinfo {title} {Quantum Phase Transitions in Transverse Field Spin Models:
				From Statistical Physics to Quantum Information}}}\ (\bibinfo  {publisher}
	{Cambridge University Press},\ \bibinfo {address} {Cambridge, UK},\ \bibinfo
	{year} {2015})\BibitemShut {NoStop}%
	\bibitem [{\citenamefont {Leggett}(2006)}]{Leggett}%
	\BibitemOpen
	\bibfield  {author} {\bibinfo {author} {\bibfnamefont {A.~J.}\ \bibnamefont
			{Leggett}},\ }\href@noop {} {{\selectlanguage {english}\emph {\bibinfo
				{title} {Quantum Liquids : Bose condensation and Cooper pairing in
					condensed-matter systems}}}},\ Oxford Graduate Texts\ (\bibinfo  {publisher}
	{Oxford University Press},\ \bibinfo {address} {Oxford, UK},\ \bibinfo {year}
	{2006})\BibitemShut {NoStop}%
	\bibitem [{\citenamefont {Zurek}\ and\ \citenamefont
		{Dorner}(2008)}]{Zurekptis}%
	\BibitemOpen
	\bibfield  {author} {\bibinfo {author} {\bibfnamefont {W.~H.}\ \bibnamefont
			{Zurek}}\ and\ \bibinfo {author} {\bibfnamefont {U.}~\bibnamefont {Dorner}},\
	}\bibfield  {title} {{\selectlanguage {english}\bibinfo {title} {Phase
				transition in space: how far does a symmetry bend before it breaks?}},\
	}\href@noop {} {\bibfield  {journal} {\bibinfo  {journal} {Philos. Trans.
				Royal Soc. A}\ }\textbf {\bibinfo {volume} {366}},\ \bibinfo {pages} {2953}
		(\bibinfo {year} {2008})}\BibitemShut {NoStop}%
	\bibitem [{\citenamefont {Dziarmaga}\ and\ \citenamefont
		{Rams}(2010)}]{Dziarmaga-doiqpt}%
	\BibitemOpen
	\bibfield  {author} {\bibinfo {author} {\bibfnamefont {J.}~\bibnamefont
			{Dziarmaga}}\ and\ \bibinfo {author} {\bibfnamefont {M.~M.}\ \bibnamefont
			{Rams}},\ }\bibfield  {title} {{\selectlanguage {english}\bibinfo {title}
			{Dynamics of an inhomogeneous quantum phase transition}},\ }\href@noop {}
	{\bibfield  {journal} {\bibinfo  {journal} {New J. Phys.}\ }\textbf {\bibinfo
			{volume} {12}},\ \bibinfo {pages} {055007} (\bibinfo {year}
		{2010})}\BibitemShut {NoStop}%
	\bibitem [{\citenamefont {Damski}\ and\ \citenamefont
		{Zurek}(2009)}]{Damski_2009}%
	\BibitemOpen
	\bibfield  {author} {\bibinfo {author} {\bibfnamefont {B.}~\bibnamefont
			{Damski}}\ and\ \bibinfo {author} {\bibfnamefont {W.~H.}\ \bibnamefont
			{Zurek}},\ }\bibfield  {title} {\bibinfo {title} {Quantum phase transition in
			space in a ferromagnetic spin-1 bose–einstein condensate},\ }\href
	{https://doi.org/10.1088/1367-2630/11/6/063014} {\bibfield  {journal}
		{\bibinfo  {journal} {New J. Phys.}\ }\textbf {\bibinfo {volume} {11}},\
		\bibinfo {pages} {063014} (\bibinfo {year} {2009})}\BibitemShut {NoStop}%
	\bibitem [{\citenamefont {Dziarmaga}(2010)}]{Dziarmaga-rev}%
	\BibitemOpen
	\bibfield  {author} {\bibinfo {author} {\bibfnamefont {J.}~\bibnamefont
			{Dziarmaga}},\ }\bibfield  {title} {{\selectlanguage {english}\bibinfo
			{title} {Dynamics of a quantum phase transition and relaxation to a steady
				state}},\ }\href@noop {} {\bibfield  {journal} {\bibinfo  {journal} {Adv.
				Phys.}\ }\textbf {\bibinfo {volume} {59}},\ \bibinfo {pages} {1063} (\bibinfo
		{year} {2010})}\BibitemShut {NoStop}%
	\bibitem [{\citenamefont {Lacki}\ and\ \citenamefont {Damski}(2017)}]{Lacki17}%
	\BibitemOpen
	\bibfield  {author} {\bibinfo {author} {\bibfnamefont {M.}~\bibnamefont
			{Lacki}}\ and\ \bibinfo {author} {\bibfnamefont {B.}~\bibnamefont {Damski}},\
	}\bibfield  {title} {\bibinfo {title} {Spatial kibble–zurek mechanism
			through susceptibilities: the inhomogeneous quantum ising model case},\
	}\href {https://doi.org/10.1088/1742-5468/aa8c20} {\bibfield  {journal}
		{\bibinfo  {journal} {J. Stat. Mech.}\ ,\ \bibinfo {pages} {103105}}
		(\bibinfo {year} {2017})}\BibitemShut {NoStop}%
	\bibitem [{\citenamefont {Parajuli}\ and\ \citenamefont
		{Chien}(2023)}]{Parajuli23}%
	\BibitemOpen
	\bibfield  {author} {\bibinfo {author} {\bibfnamefont {B.}~\bibnamefont
			{Parajuli}}\ and\ \bibinfo {author} {\bibfnamefont {C.-C.}\ \bibnamefont
			{Chien}},\ }\bibfield  {title} {\bibinfo {title} {Proximity effect and
			spatial kibble-zurek mechanism in atomic fermi gases with inhomogeneous
			pairing interactions},\ }\href {https://doi.org/10.1103/PhysRevA.107.063314}
	{\bibfield  {journal} {\bibinfo  {journal} {Phys. Rev. A}\ }\textbf {\bibinfo
			{volume} {107}},\ \bibinfo {pages} {063314} (\bibinfo {year}
		{2023})}\BibitemShut {NoStop}%
	\bibitem [{\citenamefont {Kitaev}(2001)}]{Kitaev-chain}%
	\BibitemOpen
	\bibfield  {author} {\bibinfo {author} {\bibfnamefont {A.~Y.}\ \bibnamefont
			{Kitaev}},\ }\bibfield  {title} {\bibinfo {title} {Unpaired majorana fermions
			in quantum wires},\ }\href {https://doi.org/10.1070/1063-7869/44/10S/S29}
	{\bibfield  {journal} {\bibinfo  {journal} {Phys. Usp.}\ }\textbf {\bibinfo
			{volume} {44}},\ \bibinfo {pages} {131} (\bibinfo {year} {2001})}\BibitemShut
	{NoStop}%
	\bibitem [{\citenamefont {Takagi}\ \emph {et~al.}(2020)\citenamefont {Takagi},
		\citenamefont {Tamura},\ and\ \citenamefont {Tanaka}}]{PhysRevB.101.024509}%
	\BibitemOpen
	\bibfield  {author} {\bibinfo {author} {\bibfnamefont {D.}~\bibnamefont
			{Takagi}}, \bibinfo {author} {\bibfnamefont {S.}~\bibnamefont {Tamura}},\
		and\ \bibinfo {author} {\bibfnamefont {Y.}~\bibnamefont {Tanaka}},\
	}\bibfield  {title} {\bibinfo {title} {Odd-frequency pairing and proximity
			effect in kitaev chain systems including a topological critical point},\
	}\href {https://doi.org/10.1103/PhysRevB.101.024509} {\bibfield  {journal}
		{\bibinfo  {journal} {Phys. Rev. B}\ }\textbf {\bibinfo {volume} {101}},\
		\bibinfo {pages} {024509} (\bibinfo {year} {2020})}\BibitemShut {NoStop}%
	\bibitem [{\citenamefont {De~Gennes}(2018)}]{degennes-sc}%
	\BibitemOpen
	\bibfield  {author} {\bibinfo {author} {\bibfnamefont {P.~G.}\ \bibnamefont
			{De~Gennes}},\ }\href@noop {} {{\selectlanguage {english}\emph {\bibinfo
				{title} {Superconductivity of Metals and Alloys.}}}},\ \bibinfo {edition}
	{2nd}\ ed.,\ Advanced Books Classics\ (\bibinfo  {publisher} {Chapman and
		Hall/CRC},\ \bibinfo {address} {Boulder},\ \bibinfo {year}
	{2018})\BibitemShut {NoStop}%
	\bibitem [{\citenamefont {Zhu}(2016)}]{BdG-book}%
	\BibitemOpen
	\bibfield  {author} {\bibinfo {author} {\bibfnamefont {J.-X.}\ \bibnamefont
			{Zhu}},\ }\href@noop {} {{\selectlanguage {english}\emph {\bibinfo {title}
				{Bogoliubov-de Gennes Method and Its Applications}}}},\ \bibinfo {edition}
	{1st}\ ed.,\ Lecture Notes in Physics, 924\ (\bibinfo  {publisher} {Springer
		International Publishing},\ \bibinfo {address} {Cham},\ \bibinfo {year}
	{2016})\BibitemShut {NoStop}%
	\bibitem [{\citenamefont {Hasan}\ and\ \citenamefont
		{Kane}(2010)}]{Kane_TIRev}%
	\BibitemOpen
	\bibfield  {author} {\bibinfo {author} {\bibfnamefont {M.~Z.}\ \bibnamefont
			{Hasan}}\ and\ \bibinfo {author} {\bibfnamefont {C.~L.}\ \bibnamefont
			{Kane}},\ }\bibfield  {title} {\bibinfo {title} {Colloquium: Topological
			insulators},\ }\href@noop {} {\bibfield  {journal} {\bibinfo  {journal} {Rev.
				Mod. Phys.}\ }\textbf {\bibinfo {volume} {82}},\ \bibinfo {pages} {3045}
		(\bibinfo {year} {2010})}\BibitemShut {NoStop}%
	\bibitem [{\citenamefont {Qi}\ and\ \citenamefont {Zhang}(2011)}]{Zhang_TIRev}%
	\BibitemOpen
	\bibfield  {author} {\bibinfo {author} {\bibfnamefont {X.-L.}\ \bibnamefont
			{Qi}}\ and\ \bibinfo {author} {\bibfnamefont {S.-C.}\ \bibnamefont {Zhang}},\
	}\bibfield  {title} {\bibinfo {title} {Topological insulators and
			superconductors},\ }\href@noop {} {\bibfield  {journal} {\bibinfo  {journal}
			{Rev. Mod. Phys.}\ }\textbf {\bibinfo {volume} {83}},\ \bibinfo {pages}
		{1057} (\bibinfo {year} {2011})}\BibitemShut {NoStop}%
	\bibitem [{\citenamefont {Chiu}\ \emph {et~al.}(2016)\citenamefont {Chiu},
		\citenamefont {Teo}, \citenamefont {Schnyder},\ and\ \citenamefont
		{Ryu}}]{ChiuRMP}%
	\BibitemOpen
	\bibfield  {author} {\bibinfo {author} {\bibfnamefont {C.~K.}\ \bibnamefont
			{Chiu}}, \bibinfo {author} {\bibfnamefont {J.~C.~Y.}\ \bibnamefont {Teo}},
		\bibinfo {author} {\bibfnamefont {A.~P.}\ \bibnamefont {Schnyder}},\ and\
		\bibinfo {author} {\bibfnamefont {S.}~\bibnamefont {Ryu}},\ }\bibfield
	{title} {\bibinfo {title} {Classification of topological quantum matter with
			symmetries},\ }\href@noop {} {\bibfield  {journal} {\bibinfo  {journal} {Rev.
				Mod. Phys.}\ }\textbf {\bibinfo {volume} {88}},\ \bibinfo {pages} {035005}
		(\bibinfo {year} {2016})}\BibitemShut {NoStop}%
	\bibitem [{\citenamefont {Fu}\ and\ \citenamefont
		{Kane}(2008)}]{PhysRevLett.100.096407}%
	\BibitemOpen
	\bibfield  {author} {\bibinfo {author} {\bibfnamefont {L.}~\bibnamefont
			{Fu}}\ and\ \bibinfo {author} {\bibfnamefont {C.~L.}\ \bibnamefont {Kane}},\
	}\bibfield  {title} {\bibinfo {title} {Superconducting proximity effect and
			majorana fermions at the surface of a topological insulator},\ }\href
	{https://doi.org/10.1103/PhysRevLett.100.096407} {\bibfield  {journal}
		{\bibinfo  {journal} {Phys. Rev. Lett.}\ }\textbf {\bibinfo {volume} {100}},\
		\bibinfo {pages} {096407} (\bibinfo {year} {2008})}\BibitemShut {NoStop}%
	\bibitem [{\citenamefont {Klinovaja}\ and\ \citenamefont
		{Loss}(2012)}]{PhysRevB.86.085408}%
	\BibitemOpen
	\bibfield  {author} {\bibinfo {author} {\bibfnamefont {J.}~\bibnamefont
			{Klinovaja}}\ and\ \bibinfo {author} {\bibfnamefont {D.}~\bibnamefont
			{Loss}},\ }\bibfield  {title} {\bibinfo {title} {Composite majorana fermion
			wave functions in nanowires},\ }\href
	{https://doi.org/10.1103/PhysRevB.86.085408} {\bibfield  {journal} {\bibinfo
			{journal} {Phys. Rev. B}\ }\textbf {\bibinfo {volume} {86}},\ \bibinfo
		{pages} {085408} (\bibinfo {year} {2012})}\BibitemShut {NoStop}%
	\bibitem [{\citenamefont {Rainis}\ \emph {et~al.}(2013)\citenamefont {Rainis},
		\citenamefont {Trifunovic}, \citenamefont {Klinovaja},\ and\ \citenamefont
		{Loss}}]{PhysRevB.87.024515}%
	\BibitemOpen
	\bibfield  {author} {\bibinfo {author} {\bibfnamefont {D.}~\bibnamefont
			{Rainis}}, \bibinfo {author} {\bibfnamefont {L.}~\bibnamefont {Trifunovic}},
		\bibinfo {author} {\bibfnamefont {J.}~\bibnamefont {Klinovaja}},\ and\
		\bibinfo {author} {\bibfnamefont {D.}~\bibnamefont {Loss}},\ }\bibfield
	{title} {\bibinfo {title} {Towards a realistic transport modeling in a
			superconducting nanowire with majorana fermions},\ }\href
	{https://doi.org/10.1103/PhysRevB.87.024515} {\bibfield  {journal} {\bibinfo
			{journal} {Phys. Rev. B}\ }\textbf {\bibinfo {volume} {87}},\ \bibinfo
		{pages} {024515} (\bibinfo {year} {2013})}\BibitemShut {NoStop}%
	\bibitem [{\citenamefont {Reeg}\ \emph {et~al.}(2018)\citenamefont {Reeg},
		\citenamefont {Dmytruk}, \citenamefont {Chevallier}, \citenamefont {Loss},\
		and\ \citenamefont {Klinovaja}}]{PhysRevB.98.245407}%
	\BibitemOpen
	\bibfield  {author} {\bibinfo {author} {\bibfnamefont {C.}~\bibnamefont
			{Reeg}}, \bibinfo {author} {\bibfnamefont {O.}~\bibnamefont {Dmytruk}},
		\bibinfo {author} {\bibfnamefont {D.}~\bibnamefont {Chevallier}}, \bibinfo
		{author} {\bibfnamefont {D.}~\bibnamefont {Loss}},\ and\ \bibinfo {author}
		{\bibfnamefont {J.}~\bibnamefont {Klinovaja}},\ }\bibfield  {title} {\bibinfo
		{title} {Zero-energy andreev bound states from quantum dots in proximitized
			rashba nanowires},\ }\href {https://doi.org/10.1103/PhysRevB.98.245407}
	{\bibfield  {journal} {\bibinfo  {journal} {Phys. Rev. B}\ }\textbf {\bibinfo
			{volume} {98}},\ \bibinfo {pages} {245407} (\bibinfo {year}
		{2018})}\BibitemShut {NoStop}%
	\bibitem [{\citenamefont {Kibble}(1976)}]{kibble-1976}%
	\BibitemOpen
	\bibfield  {author} {\bibinfo {author} {\bibfnamefont {T.~W.~B.}\
			\bibnamefont {Kibble}},\ }\bibfield  {title} {\bibinfo {title} {Topology of
			cosmic domains and strings},\ }\href
	{https://doi.org/10.1088/0305-4470/9/8/029} {\bibfield  {journal} {\bibinfo
			{journal} {J. Phys. A: Math. Gen.}\ }\textbf {\bibinfo {volume} {9}},\
		\bibinfo {pages} {1387} (\bibinfo {year} {1976})}\BibitemShut {NoStop}%
	\bibitem [{\citenamefont {Kibble}(1980)}]{kibble-1980}%
	\BibitemOpen
	\bibfield  {author} {\bibinfo {author} {\bibfnamefont {T.}~\bibnamefont
			{Kibble}},\ }\bibfield  {title} {\bibinfo {title} {Some implications of a
			cosmological phase transition},\ }\href
	{https://doi.org/https://doi.org/10.1016/0370-1573(80)90091-5} {\bibfield
		{journal} {\bibinfo  {journal} {Phys. Rep.}\ }\textbf {\bibinfo {volume}
			{67}},\ \bibinfo {pages} {183} (\bibinfo {year} {1980})}\BibitemShut
	{NoStop}%
	\bibitem [{\citenamefont {Zurek}(1985)}]{Zurek-1985}%
	\BibitemOpen
	\bibfield  {author} {\bibinfo {author} {\bibfnamefont {W.~H.}\ \bibnamefont
			{Zurek}},\ }\bibfield  {title} {{\selectlanguage {english}\bibinfo {title}
			{Cosmological experiments in superfluid helium?}},\ }\href@noop {} {\bibfield
		{journal} {\bibinfo  {journal} {Nature (London)}\ }\textbf {\bibinfo
			{volume} {317}},\ \bibinfo {pages} {505} (\bibinfo {year}
		{1985})}\BibitemShut {NoStop}%
	\bibitem [{\citenamefont {Zurek}(1996)}]{Zurek-1996}%
	\BibitemOpen
	\bibfield  {author} {\bibinfo {author} {\bibfnamefont {W.}~\bibnamefont
			{Zurek}},\ }\bibfield  {title} {\bibinfo {title} {Cosmological experiments in
			condensed matter systems},\ }\href
	{https://doi.org/10.1016/s0370-1573(96)00009-9} {\bibfield  {journal}
		{\bibinfo  {journal} {Phys. Rep.}\ }\textbf {\bibinfo {volume} {276}},\
		\bibinfo {pages} {177} (\bibinfo {year} {1996})}\BibitemShut {NoStop}%
	\bibitem [{\citenamefont {Laguna}\ and\ \citenamefont
		{Zurek}(1997)}]{Zurek-1997}%
	\BibitemOpen
	\bibfield  {author} {\bibinfo {author} {\bibfnamefont {P.}~\bibnamefont
			{Laguna}}\ and\ \bibinfo {author} {\bibfnamefont {W.~H.}\ \bibnamefont
			{Zurek}},\ }\bibfield  {title} {\bibinfo {title} {Density of kinks after a
			quench: When symmetry breaks, how big are the pieces?},\ }\href
	{https://doi.org/10.1103/PhysRevLett.78.2519} {\bibfield  {journal} {\bibinfo
			{journal} {Phys. Rev. Lett.}\ }\textbf {\bibinfo {volume} {78}},\ \bibinfo
		{pages} {2519} (\bibinfo {year} {1997})}\BibitemShut {NoStop}%
	\bibitem [{\citenamefont {Anglin}\ and\ \citenamefont
		{Zurek}(1999)}]{Zurek-1999}%
	\BibitemOpen
	\bibfield  {author} {\bibinfo {author} {\bibfnamefont {J.~R.}\ \bibnamefont
			{Anglin}}\ and\ \bibinfo {author} {\bibfnamefont {W.~H.}\ \bibnamefont
			{Zurek}},\ }\bibfield  {title} {\bibinfo {title} {Vortices in the wake of
			rapid bose-einstein condensation},\ }\href
	{https://doi.org/10.1103/PhysRevLett.83.1707} {\bibfield  {journal} {\bibinfo
			{journal} {Phys. Rev. Lett.}\ }\textbf {\bibinfo {volume} {83}},\ \bibinfo
		{pages} {1707} (\bibinfo {year} {1999})}\BibitemShut {NoStop}%
	\bibitem [{\citenamefont {Stephens}\ \emph {et~al.}(2002)\citenamefont
		{Stephens}, \citenamefont {Bettencourt},\ and\ \citenamefont
		{Zurek}}]{Zurek-2002}%
	\BibitemOpen
	\bibfield  {author} {\bibinfo {author} {\bibfnamefont {G.~J.}\ \bibnamefont
			{Stephens}}, \bibinfo {author} {\bibfnamefont {L.~M.~A.}\ \bibnamefont
			{Bettencourt}},\ and\ \bibinfo {author} {\bibfnamefont {W.~H.}\ \bibnamefont
			{Zurek}},\ }\bibfield  {title} {\bibinfo {title} {Critical dynamics of gauge
			systems: Spontaneous vortex formation in 2d superconductors},\ }\href
	{https://doi.org/10.1103/PhysRevLett.88.137004} {\bibfield  {journal}
		{\bibinfo  {journal} {Phys. Rev. Lett.}\ }\textbf {\bibinfo {volume} {88}},\
		\bibinfo {pages} {137004} (\bibinfo {year} {2002})}\BibitemShut {NoStop}%
	\bibitem [{\citenamefont {Dziarmaga}(2005)}]{QPT-Dziarmaga-2005}%
	\BibitemOpen
	\bibfield  {author} {\bibinfo {author} {\bibfnamefont {J.}~\bibnamefont
			{Dziarmaga}},\ }\bibfield  {title} {\bibinfo {title} {Dynamics of a quantum
			phase transition: Exact solution of the quantum ising model},\ }\href
	{https://doi.org/10.1103/PhysRevLett.95.245701} {\bibfield  {journal}
		{\bibinfo  {journal} {Phys. Rev. Lett.}\ }\textbf {\bibinfo {volume} {95}},\
		\bibinfo {pages} {245701} (\bibinfo {year} {2005})}\BibitemShut {NoStop}%
	\bibitem [{\citenamefont {Zurek}\ \emph {et~al.}(2005)\citenamefont {Zurek},
		\citenamefont {Dorner},\ and\ \citenamefont {Zoller}}]{QPT-Zurek-2005}%
	\BibitemOpen
	\bibfield  {author} {\bibinfo {author} {\bibfnamefont {W.~H.}\ \bibnamefont
			{Zurek}}, \bibinfo {author} {\bibfnamefont {U.}~\bibnamefont {Dorner}},\ and\
		\bibinfo {author} {\bibfnamefont {P.}~\bibnamefont {Zoller}},\ }\bibfield
	{title} {\bibinfo {title} {Dynamics of a quantum phase transition},\ }\href
	{https://doi.org/10.1103/PhysRevLett.95.105701} {\bibfield  {journal}
		{\bibinfo  {journal} {Phys. Rev. Lett.}\ }\textbf {\bibinfo {volume} {95}},\
		\bibinfo {pages} {105701} (\bibinfo {year} {2005})}\BibitemShut {NoStop}%
	\bibitem [{\citenamefont {Jaschke}\ \emph {et~al.}(2017)\citenamefont
		{Jaschke}, \citenamefont {Maeda}, \citenamefont {Whalen}, \citenamefont
		{Wall},\ and\ \citenamefont {Carr}}]{QIM-Jaschke-2017}%
	\BibitemOpen
	\bibfield  {author} {\bibinfo {author} {\bibfnamefont {D.}~\bibnamefont
			{Jaschke}}, \bibinfo {author} {\bibfnamefont {K.}~\bibnamefont {Maeda}},
		\bibinfo {author} {\bibfnamefont {J.~D.}\ \bibnamefont {Whalen}}, \bibinfo
		{author} {\bibfnamefont {M.~L.}\ \bibnamefont {Wall}},\ and\ \bibinfo
		{author} {\bibfnamefont {L.~D.}\ \bibnamefont {Carr}},\ }\bibfield  {title}
	{\bibinfo {title} {Critical phenomena and kibble–zurek scaling in the
			long-range quantum ising chain},\ }\href
	{https://doi.org/10.1088/1367-2630/aa65bc} {\bibfield  {journal} {\bibinfo
			{journal} {New J. Phys.}\ }\textbf {\bibinfo {volume} {19}},\ \bibinfo
		{pages} {033032} (\bibinfo {year} {2017})}\BibitemShut {NoStop}%
	\bibitem [{\citenamefont {Dziarmaga}\ \emph {et~al.}(2022)\citenamefont
		{Dziarmaga}, \citenamefont {Rams},\ and\ \citenamefont
		{Zurek}}]{Dziarmaga-2022}%
	\BibitemOpen
	\bibfield  {author} {\bibinfo {author} {\bibfnamefont {J.}~\bibnamefont
			{Dziarmaga}}, \bibinfo {author} {\bibfnamefont {M.~M.}\ \bibnamefont
			{Rams}},\ and\ \bibinfo {author} {\bibfnamefont {W.~H.}\ \bibnamefont
			{Zurek}},\ }\bibfield  {title} {\bibinfo {title} {Coherent many-body
			oscillations induced by a superposition of broken symmetry states in the wake
			of a quantum phase transition},\ }\href
	{https://doi.org/10.1103/physrevlett.129.260407} {\bibfield  {journal}
		{\bibinfo  {journal} {Phys. Rev. Lett.}\ }\textbf {\bibinfo {volume} {129}},\
		\bibinfo {pages} {260407} (\bibinfo {year} {2022})}\BibitemShut {NoStop}%
	\bibitem [{\citenamefont {Uhlmann}\ \emph
		{et~al.}(2010{\natexlab{a}})\citenamefont {Uhlmann}, \citenamefont
		{Sch\"utzhold},\ and\ \citenamefont {Fischer}}]{PhysRevD.81.025017}%
	\BibitemOpen
	\bibfield  {author} {\bibinfo {author} {\bibfnamefont {M.}~\bibnamefont
			{Uhlmann}}, \bibinfo {author} {\bibfnamefont {R.}~\bibnamefont
			{Sch\"utzhold}},\ and\ \bibinfo {author} {\bibfnamefont {U.~R.}\ \bibnamefont
			{Fischer}},\ }\bibfield  {title} {\bibinfo {title} {$o(n)$ symmetry-breaking
			quantum quench: Topological defects versus quasiparticles},\ }\href
	{https://doi.org/10.1103/PhysRevD.81.025017} {\bibfield  {journal} {\bibinfo
			{journal} {Phys. Rev. D}\ }\textbf {\bibinfo {volume} {81}},\ \bibinfo
		{pages} {025017} (\bibinfo {year} {2010}{\natexlab{a}})}\BibitemShut
	{NoStop}%
	\bibitem [{\citenamefont {Uhlmann}\ \emph
		{et~al.}(2010{\natexlab{b}})\citenamefont {Uhlmann}, \citenamefont
		{Schützhold},\ and\ \citenamefont {Fischer}}]{Uhlmann_2010}%
	\BibitemOpen
	\bibfield  {author} {\bibinfo {author} {\bibfnamefont {M.}~\bibnamefont
			{Uhlmann}}, \bibinfo {author} {\bibfnamefont {R.}~\bibnamefont
			{Schützhold}},\ and\ \bibinfo {author} {\bibfnamefont {U.~R.}\ \bibnamefont
			{Fischer}},\ }\bibfield  {title} {\bibinfo {title} {System size scaling of
			topological defect creation in a second-order dynamical quantum phase
			transition},\ }\href {https://doi.org/10.1088/1367-2630/12/9/095020}
	{\bibfield  {journal} {\bibinfo  {journal} {New J. Phys.}\ }\textbf {\bibinfo
			{volume} {12}},\ \bibinfo {pages} {095020} (\bibinfo {year}
		{2010}{\natexlab{b}})}\BibitemShut {NoStop}%
	\bibitem [{\citenamefont {Polkovnikov}(2005)}]{adiabtic-dynamics-2005}%
	\BibitemOpen
	\bibfield  {author} {\bibinfo {author} {\bibfnamefont {A.}~\bibnamefont
			{Polkovnikov}},\ }\bibfield  {title} {\bibinfo {title} {Universal adiabatic
			dynamics in the vicinity of a quantum critical point},\ }\href
	{https://doi.org/10.1103/PhysRevB.72.161201} {\bibfield  {journal} {\bibinfo
			{journal} {Phys. Rev. B}\ }\textbf {\bibinfo {volume} {72}},\ \bibinfo
		{pages} {161201(R)} (\bibinfo {year} {2005})}\BibitemShut {NoStop}%
	\bibitem [{\citenamefont {Warner}\ and\ \citenamefont
		{Leggett}(2005)}]{QD-Legget-2005}%
	\BibitemOpen
	\bibfield  {author} {\bibinfo {author} {\bibfnamefont {G.~L.}\ \bibnamefont
			{Warner}}\ and\ \bibinfo {author} {\bibfnamefont {A.~J.}\ \bibnamefont
			{Leggett}},\ }\bibfield  {title} {\bibinfo {title} {Quench dynamics of a
			superfluid fermi gas},\ }\href {https://doi.org/10.1103/PhysRevB.71.134514}
	{\bibfield  {journal} {\bibinfo  {journal} {Phys. Rev. B}\ }\textbf {\bibinfo
			{volume} {71}},\ \bibinfo {pages} {134514} (\bibinfo {year}
		{2005})}\BibitemShut {NoStop}%
	\bibitem [{\citenamefont {Shimizu}\ \emph {et~al.}(2018)\citenamefont
		{Shimizu}, \citenamefont {Kuno}, \citenamefont {Hirano},\ and\ \citenamefont
		{Ichinose}}]{PhysRevA.97.033626}%
	\BibitemOpen
	\bibfield  {author} {\bibinfo {author} {\bibfnamefont {K.}~\bibnamefont
			{Shimizu}}, \bibinfo {author} {\bibfnamefont {Y.}~\bibnamefont {Kuno}},
		\bibinfo {author} {\bibfnamefont {T.}~\bibnamefont {Hirano}},\ and\ \bibinfo
		{author} {\bibfnamefont {I.}~\bibnamefont {Ichinose}},\ }\bibfield  {title}
	{\bibinfo {title} {Dynamics of a quantum phase transition in the bose-hubbard
			model: Kibble-zurek mechanism and beyond},\ }\href
	{https://doi.org/10.1103/PhysRevA.97.033626} {\bibfield  {journal} {\bibinfo
			{journal} {Phys. Rev. A}\ }\textbf {\bibinfo {volume} {97}},\ \bibinfo
		{pages} {033626} (\bibinfo {year} {2018})}\BibitemShut {NoStop}%
	\bibitem [{\citenamefont {Cucchietti}\ \emph {et~al.}(2007)\citenamefont
		{Cucchietti}, \citenamefont {Damski}, \citenamefont {Dziarmaga},\ and\
		\citenamefont {Zurek}}]{PhysRevA.75.023603}%
	\BibitemOpen
	\bibfield  {author} {\bibinfo {author} {\bibfnamefont {F.~M.}\ \bibnamefont
			{Cucchietti}}, \bibinfo {author} {\bibfnamefont {B.}~\bibnamefont {Damski}},
		\bibinfo {author} {\bibfnamefont {J.}~\bibnamefont {Dziarmaga}},\ and\
		\bibinfo {author} {\bibfnamefont {W.~H.}\ \bibnamefont {Zurek}},\ }\bibfield
	{title} {\bibinfo {title} {Dynamics of the bose-hubbard model: Transition
			from a mott insulator to a superfluid},\ }\href
	{https://doi.org/10.1103/PhysRevA.75.023603} {\bibfield  {journal} {\bibinfo
			{journal} {Phys. Rev. A}\ }\textbf {\bibinfo {volume} {75}},\ \bibinfo
		{pages} {023603} (\bibinfo {year} {2007})}\BibitemShut {NoStop}%
	\bibitem [{\citenamefont {Dziarmaga}\ \emph {et~al.}(2012)\citenamefont
		{Dziarmaga}, \citenamefont {Tylutki},\ and\ \citenamefont
		{Zurek}}]{PhysRevB.86.144521}%
	\BibitemOpen
	\bibfield  {author} {\bibinfo {author} {\bibfnamefont {J.}~\bibnamefont
			{Dziarmaga}}, \bibinfo {author} {\bibfnamefont {M.}~\bibnamefont {Tylutki}},\
		and\ \bibinfo {author} {\bibfnamefont {W.~H.}\ \bibnamefont {Zurek}},\
	}\bibfield  {title} {\bibinfo {title} {Quench from mott insulator to
			superfluid},\ }\href {https://doi.org/10.1103/PhysRevB.86.144521} {\bibfield
		{journal} {\bibinfo  {journal} {Phys. Rev. B}\ }\textbf {\bibinfo {volume}
			{86}},\ \bibinfo {pages} {144521} (\bibinfo {year} {2012})}\BibitemShut
	{NoStop}%
	\bibitem [{\citenamefont {Gardas}\ \emph {et~al.}(2017)\citenamefont {Gardas},
		\citenamefont {Dziarmaga},\ and\ \citenamefont {Zurek}}]{PhysRevB.95.104306}%
	\BibitemOpen
	\bibfield  {author} {\bibinfo {author} {\bibfnamefont {B.}~\bibnamefont
			{Gardas}}, \bibinfo {author} {\bibfnamefont {J.}~\bibnamefont {Dziarmaga}},\
		and\ \bibinfo {author} {\bibfnamefont {W.~H.}\ \bibnamefont {Zurek}},\
	}\bibfield  {title} {\bibinfo {title} {Dynamics of the quantum phase
			transition in the one-dimensional bose-hubbard model: Excitations and
			correlations induced by a quench},\ }\href
	{https://doi.org/10.1103/PhysRevB.95.104306} {\bibfield  {journal} {\bibinfo
			{journal} {Phys. Rev. B}\ }\textbf {\bibinfo {volume} {95}},\ \bibinfo
		{pages} {104306} (\bibinfo {year} {2017})}\BibitemShut {NoStop}%
	\bibitem [{\citenamefont {Machida}\ and\ \citenamefont
		{Kasamatsu}(2021)}]{PhysRevA.103.013310}%
	\BibitemOpen
	\bibfield  {author} {\bibinfo {author} {\bibfnamefont {Y.}~\bibnamefont
			{Machida}}\ and\ \bibinfo {author} {\bibfnamefont {K.}~\bibnamefont
			{Kasamatsu}},\ }\bibfield  {title} {\bibinfo {title} {Application of the
			inhomogeneous kibble-zurek mechanism to quench dynamics in the transition
			from a mott insulator to a superfluid in a finite system},\ }\href
	{https://doi.org/10.1103/PhysRevA.103.013310} {\bibfield  {journal} {\bibinfo
			{journal} {Phys. Rev. A}\ }\textbf {\bibinfo {volume} {103}},\ \bibinfo
		{pages} {013310} (\bibinfo {year} {2021})}\BibitemShut {NoStop}%
	\bibitem [{\citenamefont {Sim}\ \emph {et~al.}(2022)\citenamefont {Sim},
		\citenamefont {Chitra},\ and\ \citenamefont
		{Molignini}}]{PhysRevB.106.224302}%
	\BibitemOpen
	\bibfield  {author} {\bibinfo {author} {\bibfnamefont {K.}~\bibnamefont
			{Sim}}, \bibinfo {author} {\bibfnamefont {R.}~\bibnamefont {Chitra}},\ and\
		\bibinfo {author} {\bibfnamefont {P.}~\bibnamefont {Molignini}},\ }\bibfield
	{title} {\bibinfo {title} {Quench dynamics and scaling laws in topological
			nodal loop semimetals},\ }\href {https://doi.org/10.1103/PhysRevB.106.224302}
	{\bibfield  {journal} {\bibinfo  {journal} {Phys. Rev. B}\ }\textbf {\bibinfo
			{volume} {106}},\ \bibinfo {pages} {224302} (\bibinfo {year}
		{2022})}\BibitemShut {NoStop}%
	\bibitem [{\citenamefont {Monaco}\ \emph {et~al.}(2002)\citenamefont {Monaco},
		\citenamefont {Mygind},\ and\ \citenamefont {Rivers}}]{Monaco-2002}%
	\BibitemOpen
	\bibfield  {author} {\bibinfo {author} {\bibfnamefont {R.}~\bibnamefont
			{Monaco}}, \bibinfo {author} {\bibfnamefont {J.}~\bibnamefont {Mygind}},\
		and\ \bibinfo {author} {\bibfnamefont {R.~J.}\ \bibnamefont {Rivers}},\
	}\bibfield  {title} {\bibinfo {title} {Zurek-kibble domain structures: The
			dynamics of spontaneous vortex formation in annular josephson tunnel
			junctions},\ }\href {https://doi.org/10.1103/PhysRevLett.89.080603}
	{\bibfield  {journal} {\bibinfo  {journal} {Phys. Rev. Lett.}\ }\textbf
		{\bibinfo {volume} {89}},\ \bibinfo {pages} {080603} (\bibinfo {year}
		{2002})}\BibitemShut {NoStop}%
	\bibitem [{\citenamefont {Ulm}\ \emph {et~al.}(2013)\citenamefont {Ulm},
		\citenamefont {Rossnagel}, \citenamefont {Jacob}, \citenamefont {Deguenther},
		\citenamefont {Dawkins}, \citenamefont {Poschinger}, \citenamefont
		{Nigmatullin}, \citenamefont {Retzker}, \citenamefont {Plenio}, \citenamefont
		{Schmidt-Kaler},\ and\ \citenamefont {Singer}}]{Ulm-2013}%
	\BibitemOpen
	\bibfield  {author} {\bibinfo {author} {\bibfnamefont {S.}~\bibnamefont
			{Ulm}}, \bibinfo {author} {\bibfnamefont {J.}~\bibnamefont {Rossnagel}},
		\bibinfo {author} {\bibfnamefont {G.}~\bibnamefont {Jacob}}, \bibinfo
		{author} {\bibfnamefont {C.}~\bibnamefont {Deguenther}}, \bibinfo {author}
		{\bibfnamefont {S.~T.}\ \bibnamefont {Dawkins}}, \bibinfo {author}
		{\bibfnamefont {U.~G.}\ \bibnamefont {Poschinger}}, \bibinfo {author}
		{\bibfnamefont {R.}~\bibnamefont {Nigmatullin}}, \bibinfo {author}
		{\bibfnamefont {A.}~\bibnamefont {Retzker}}, \bibinfo {author} {\bibfnamefont
			{M.~B.}\ \bibnamefont {Plenio}}, \bibinfo {author} {\bibfnamefont
			{F.}~\bibnamefont {Schmidt-Kaler}},\ and\ \bibinfo {author} {\bibfnamefont
			{K.}~\bibnamefont {Singer}},\ }\bibfield  {title} {{\selectlanguage
			{english}\bibinfo {title} {Observation of the kibble-zurek scaling law for
				defect formation in ion crystals}},\ }\href@noop {} {\bibfield  {journal}
		{\bibinfo  {journal} {Nat. Commun.}\ }\textbf {\bibinfo {volume} {4}},\
		\bibinfo {pages} {2290} (\bibinfo {year} {2013})}\BibitemShut {NoStop}%
	\bibitem [{\citenamefont {Pyka}\ \emph {et~al.}(2013)\citenamefont {Pyka},
		\citenamefont {Keller}, \citenamefont {Partner}, \citenamefont {Nigmatullin},
		\citenamefont {Burgermeister}, \citenamefont {Meier}, \citenamefont
		{Kuhlmann}, \citenamefont {Retzker}, \citenamefont {Plenio}, \citenamefont
		{Zurek}, \citenamefont {del Campo},\ and\ \citenamefont
		{Mehlstaeubler}}]{Pyka-2013}%
	\BibitemOpen
	\bibfield  {author} {\bibinfo {author} {\bibfnamefont {K.}~\bibnamefont
			{Pyka}}, \bibinfo {author} {\bibfnamefont {J.}~\bibnamefont {Keller}},
		\bibinfo {author} {\bibfnamefont {H.~L.}\ \bibnamefont {Partner}}, \bibinfo
		{author} {\bibfnamefont {R.}~\bibnamefont {Nigmatullin}}, \bibinfo {author}
		{\bibfnamefont {T.}~\bibnamefont {Burgermeister}}, \bibinfo {author}
		{\bibfnamefont {D.~M.}\ \bibnamefont {Meier}}, \bibinfo {author}
		{\bibfnamefont {K.}~\bibnamefont {Kuhlmann}}, \bibinfo {author}
		{\bibfnamefont {A.}~\bibnamefont {Retzker}}, \bibinfo {author} {\bibfnamefont
			{M.~B.}\ \bibnamefont {Plenio}}, \bibinfo {author} {\bibfnamefont {W.~H.}\
			\bibnamefont {Zurek}}, \bibinfo {author} {\bibfnamefont {A.}~\bibnamefont
			{del Campo}},\ and\ \bibinfo {author} {\bibfnamefont {T.~E.}\ \bibnamefont
			{Mehlstaeubler}},\ }\bibfield  {title} {{\selectlanguage {english}\bibinfo
			{title} {Topological defect formation and spontaneous symmetry breaking in
				ion coulomb crystals}},\ }\href@noop {} {\bibfield  {journal} {\bibinfo
			{journal} {Nat. Commun.}\ }\textbf {\bibinfo {volume} {4}},\ \bibinfo {pages}
		{2291} (\bibinfo {year} {2013})}\BibitemShut {NoStop}%
	\bibitem [{\citenamefont {Navon}\ \emph {et~al.}(2015)\citenamefont {Navon},
		\citenamefont {Gaunt}, \citenamefont {Smith},\ and\ \citenamefont
		{Hadzibabic}}]{ALGaunt-2015}%
	\BibitemOpen
	\bibfield  {author} {\bibinfo {author} {\bibfnamefont {N.}~\bibnamefont
			{Navon}}, \bibinfo {author} {\bibfnamefont {A.~L.}\ \bibnamefont {Gaunt}},
		\bibinfo {author} {\bibfnamefont {R.~P.}\ \bibnamefont {Smith}},\ and\
		\bibinfo {author} {\bibfnamefont {Z.}~\bibnamefont {Hadzibabic}},\ }\bibfield
	{title} {\bibinfo {title} {Critical dynamics of spontaneous symmetry
			breaking in a homogeneous bose gas},\ }\href
	{https://doi.org/10.1126/science.1258676} {\bibfield  {journal} {\bibinfo
			{journal} {Science}\ }\textbf {\bibinfo {volume} {347}},\ \bibinfo {pages}
		{167} (\bibinfo {year} {2015})},\ \Eprint
	{https://arxiv.org/abs/https://www.science.org/doi/pdf/10.1126/science.1258676}
	{https://www.science.org/doi/pdf/10.1126/science.1258676} \BibitemShut
	{NoStop}%
	\bibitem [{\citenamefont {Braun}\ \emph {et~al.}(2015)\citenamefont {Braun},
		\citenamefont {Friesdorf}, \citenamefont {Hodgman}, \citenamefont
		{Schreiber}, \citenamefont {Ronzheimer}, \citenamefont {Riera}, \citenamefont
		{del Rey}, \citenamefont {Bloch}, \citenamefont {Eisert},\ and\ \citenamefont
		{Schneider}}]{QPT-Brawn-2015}%
	\BibitemOpen
	\bibfield  {author} {\bibinfo {author} {\bibfnamefont {S.}~\bibnamefont
			{Braun}}, \bibinfo {author} {\bibfnamefont {M.}~\bibnamefont {Friesdorf}},
		\bibinfo {author} {\bibfnamefont {S.~S.}\ \bibnamefont {Hodgman}}, \bibinfo
		{author} {\bibfnamefont {M.}~\bibnamefont {Schreiber}}, \bibinfo {author}
		{\bibfnamefont {J.~P.}\ \bibnamefont {Ronzheimer}}, \bibinfo {author}
		{\bibfnamefont {A.}~\bibnamefont {Riera}}, \bibinfo {author} {\bibfnamefont
			{M.}~\bibnamefont {del Rey}}, \bibinfo {author} {\bibfnamefont
			{I.}~\bibnamefont {Bloch}}, \bibinfo {author} {\bibfnamefont
			{J.}~\bibnamefont {Eisert}},\ and\ \bibinfo {author} {\bibfnamefont
			{U.}~\bibnamefont {Schneider}},\ }\bibfield  {title} {\bibinfo {title}
		{Emergence of coherence and the dynamics of quantum phase transitions},\
	}\href {https://doi.org/10.1073/pnas.1408861112} {\bibfield  {journal}
		{\bibinfo  {journal} {PNAS}\ }\textbf {\bibinfo {volume} {112}},\ \bibinfo
		{pages} {3641} (\bibinfo {year} {2015})},\ \Eprint
	{https://arxiv.org/abs/https://www.pnas.org/doi/pdf/10.1073/pnas.1408861112}
	{https://www.pnas.org/doi/pdf/10.1073/pnas.1408861112} \BibitemShut {NoStop}%
	\bibitem [{\citenamefont {Chen}\ \emph {et~al.}(2011)\citenamefont {Chen},
		\citenamefont {White}, \citenamefont {Borries},\ and\ \citenamefont
		{DeMarco}}]{QPT-mott-2011}%
	\BibitemOpen
	\bibfield  {author} {\bibinfo {author} {\bibfnamefont {D.}~\bibnamefont
			{Chen}}, \bibinfo {author} {\bibfnamefont {M.}~\bibnamefont {White}},
		\bibinfo {author} {\bibfnamefont {C.}~\bibnamefont {Borries}},\ and\ \bibinfo
		{author} {\bibfnamefont {B.}~\bibnamefont {DeMarco}},\ }\bibfield  {title}
	{\bibinfo {title} {Quantum quench of an atomic mott insulator},\ }\href
	{https://doi.org/10.1103/PhysRevLett.106.235304} {\bibfield  {journal}
		{\bibinfo  {journal} {Phys. Rev. Lett.}\ }\textbf {\bibinfo {volume} {106}},\
		\bibinfo {pages} {235304} (\bibinfo {year} {2011})}\BibitemShut {NoStop}%
	\bibitem [{\citenamefont {Keesling}\ \emph {et~al.}(2019)\citenamefont
		{Keesling}, \citenamefont {Omran}, \citenamefont {Levine}, \citenamefont
		{Bernien}, \citenamefont {Pichler}, \citenamefont {Choi}, \citenamefont
		{Samajdar}, \citenamefont {Schwartz}, \citenamefont {Silvi}, \citenamefont
		{Sachdev}, \citenamefont {Zoller}, \citenamefont {Endres}, \citenamefont
		{Greiner}, \citenamefont {Vuletic},\ and\ \citenamefont
		{Lukin}}]{QKZM-Nature-2019}%
	\BibitemOpen
	\bibfield  {author} {\bibinfo {author} {\bibfnamefont {A.}~\bibnamefont
			{Keesling}}, \bibinfo {author} {\bibfnamefont {A.}~\bibnamefont {Omran}},
		\bibinfo {author} {\bibfnamefont {H.}~\bibnamefont {Levine}}, \bibinfo
		{author} {\bibfnamefont {H.}~\bibnamefont {Bernien}}, \bibinfo {author}
		{\bibfnamefont {H.}~\bibnamefont {Pichler}}, \bibinfo {author} {\bibfnamefont
			{S.}~\bibnamefont {Choi}}, \bibinfo {author} {\bibfnamefont {R.}~\bibnamefont
			{Samajdar}}, \bibinfo {author} {\bibfnamefont {S.}~\bibnamefont {Schwartz}},
		\bibinfo {author} {\bibfnamefont {P.}~\bibnamefont {Silvi}}, \bibinfo
		{author} {\bibfnamefont {S.}~\bibnamefont {Sachdev}}, \bibinfo {author}
		{\bibfnamefont {P.}~\bibnamefont {Zoller}}, \bibinfo {author} {\bibfnamefont
			{M.}~\bibnamefont {Endres}}, \bibinfo {author} {\bibfnamefont
			{M.}~\bibnamefont {Greiner}}, \bibinfo {author} {\bibfnamefont
			{V.}~\bibnamefont {Vuletic}},\ and\ \bibinfo {author} {\bibfnamefont {M.~D.}\
			\bibnamefont {Lukin}},\ }\bibfield  {title} {{\selectlanguage
			{english}\bibinfo {title} {Quantum kibble-zurek mechanism and critical
				dynamics on a programmable rydberg simulator}},\ }\href@noop {} {\bibfield
		{journal} {\bibinfo  {journal} {Nature}\ }\textbf {\bibinfo {volume} {568}},\
		\bibinfo {pages} {207} (\bibinfo {year} {2019})}\BibitemShut {NoStop}%
	\bibitem [{\citenamefont {Anquez}\ \emph {et~al.}(2016)\citenamefont {Anquez},
		\citenamefont {Robbins}, \citenamefont {Bharath}, \citenamefont
		{Boguslawski}, \citenamefont {Hoang},\ and\ \citenamefont
		{Chapman}}]{QKZM-prl-2016}%
	\BibitemOpen
	\bibfield  {author} {\bibinfo {author} {\bibfnamefont {M.}~\bibnamefont
			{Anquez}}, \bibinfo {author} {\bibfnamefont {B.~A.}\ \bibnamefont {Robbins}},
		\bibinfo {author} {\bibfnamefont {H.~M.}\ \bibnamefont {Bharath}}, \bibinfo
		{author} {\bibfnamefont {M.}~\bibnamefont {Boguslawski}}, \bibinfo {author}
		{\bibfnamefont {T.~M.}\ \bibnamefont {Hoang}},\ and\ \bibinfo {author}
		{\bibfnamefont {M.~S.}\ \bibnamefont {Chapman}},\ }\bibfield  {title}
	{\bibinfo {title} {Quantum kibble-zurek mechanism in a spin-1 bose-einstein
			condensate},\ }\href {https://doi.org/10.1103/PhysRevLett.116.155301}
	{\bibfield  {journal} {\bibinfo  {journal} {Phys. Rev. Lett.}\ }\textbf
		{\bibinfo {volume} {116}},\ \bibinfo {pages} {155301} (\bibinfo {year}
		{2016})}\BibitemShut {NoStop}%
	\bibitem [{\citenamefont {Li}\ \emph {et~al.}(2023)\citenamefont {Li},
		\citenamefont {Wu}, \citenamefont {Mei}, \citenamefont {Yao}, \citenamefont
		{Lian}, \citenamefont {Cai}, \citenamefont {Wang}, \citenamefont {Qi},
		\citenamefont {Yao}, \citenamefont {He}, \citenamefont {Zhou},\ and\
		\citenamefont {Duan}}]{PRXQuantum.4.010302}%
	\BibitemOpen
	\bibfield  {author} {\bibinfo {author} {\bibfnamefont {B.-W.}\ \bibnamefont
			{Li}}, \bibinfo {author} {\bibfnamefont {Y.-K.}\ \bibnamefont {Wu}}, \bibinfo
		{author} {\bibfnamefont {Q.-X.}\ \bibnamefont {Mei}}, \bibinfo {author}
		{\bibfnamefont {R.}~\bibnamefont {Yao}}, \bibinfo {author} {\bibfnamefont
			{W.-Q.}\ \bibnamefont {Lian}}, \bibinfo {author} {\bibfnamefont {M.-L.}\
			\bibnamefont {Cai}}, \bibinfo {author} {\bibfnamefont {Y.}~\bibnamefont
			{Wang}}, \bibinfo {author} {\bibfnamefont {B.-X.}\ \bibnamefont {Qi}},
		\bibinfo {author} {\bibfnamefont {L.}~\bibnamefont {Yao}}, \bibinfo {author}
		{\bibfnamefont {L.}~\bibnamefont {He}}, \bibinfo {author} {\bibfnamefont
			{Z.-C.}\ \bibnamefont {Zhou}},\ and\ \bibinfo {author} {\bibfnamefont
			{L.-M.}\ \bibnamefont {Duan}},\ }\bibfield  {title} {\bibinfo {title}
		{Probing critical behavior of long-range transverse-field ising model through
			quantum kibble-zurek mechanism},\ }\href
	{https://doi.org/10.1103/PRXQuantum.4.010302} {\bibfield  {journal} {\bibinfo
			{journal} {PRX Quantum}\ }\textbf {\bibinfo {volume} {4}},\ \bibinfo {pages}
		{010302} (\bibinfo {year} {2023})}\BibitemShut {NoStop}%
	\bibitem [{\citenamefont {Deutschländer}\ \emph {et~al.}(2015)\citenamefont
		{Deutschländer}, \citenamefont {Dillmann}, \citenamefont {Maret},\ and\
		\citenamefont {Keim}}]{KZM-colloidal-monolayer-2015}%
	\BibitemOpen
	\bibfield  {author} {\bibinfo {author} {\bibfnamefont {S.}~\bibnamefont
			{Deutschländer}}, \bibinfo {author} {\bibfnamefont {P.}~\bibnamefont
			{Dillmann}}, \bibinfo {author} {\bibfnamefont {G.}~\bibnamefont {Maret}},\
		and\ \bibinfo {author} {\bibfnamefont {P.}~\bibnamefont {Keim}},\ }\bibfield
	{title} {\bibinfo {title} {Kibble–zurek mechanism in colloidal
			monolayers},\ }\href {https://doi.org/10.1073/pnas.1500763112} {\bibfield
		{journal} {\bibinfo  {journal} {PNAS}\ }\textbf {\bibinfo {volume} {112}},\
		\bibinfo {pages} {6925} (\bibinfo {year} {2015})},\ \Eprint
	{https://arxiv.org/abs/https://www.pnas.org/doi/pdf/10.1073/pnas.1500763112}
	{https://www.pnas.org/doi/pdf/10.1073/pnas.1500763112} \BibitemShut {NoStop}%
	\bibitem [{\citenamefont {Ko}\ \emph {et~al.}(2019)\citenamefont {Ko},
		\citenamefont {Park},\ and\ \citenamefont {Shin}}]{KZM-FermiSF-2019}%
	\BibitemOpen
	\bibfield  {author} {\bibinfo {author} {\bibfnamefont {B.}~\bibnamefont
			{Ko}}, \bibinfo {author} {\bibfnamefont {J.~W.}\ \bibnamefont {Park}},\ and\
		\bibinfo {author} {\bibfnamefont {Y.}~\bibnamefont {Shin}},\ }\bibfield
	{title} {{\selectlanguage {english}\bibinfo {title} {Kibble-zurek
				universality in a strongly interacting fermi superfluid}},\ }\href@noop {}
	{\bibfield  {journal} {\bibinfo  {journal} {Nat. Phys.}\ }\textbf {\bibinfo
			{volume} {15}},\ \bibinfo {pages} {1227} (\bibinfo {year}
		{2019})}\BibitemShut {NoStop}%
	\bibitem [{\citenamefont {Liu}\ \emph {et~al.}(2021)\citenamefont {Liu},
		\citenamefont {Yao}, \citenamefont {Deng}, \citenamefont {Wang},
		\citenamefont {Wang}, \citenamefont {Li}, \citenamefont {Chen}, \citenamefont
		{Chen},\ and\ \citenamefont {Pan}}]{KZM-vortices-2021}%
	\BibitemOpen
	\bibfield  {author} {\bibinfo {author} {\bibfnamefont {X.-P.}\ \bibnamefont
			{Liu}}, \bibinfo {author} {\bibfnamefont {X.-C.}\ \bibnamefont {Yao}},
		\bibinfo {author} {\bibfnamefont {Y.}~\bibnamefont {Deng}}, \bibinfo {author}
		{\bibfnamefont {Y.-X.}\ \bibnamefont {Wang}}, \bibinfo {author}
		{\bibfnamefont {X.-Q.}\ \bibnamefont {Wang}}, \bibinfo {author}
		{\bibfnamefont {X.}~\bibnamefont {Li}}, \bibinfo {author} {\bibfnamefont
			{Q.}~\bibnamefont {Chen}}, \bibinfo {author} {\bibfnamefont {Y.-A.}\
			\bibnamefont {Chen}},\ and\ \bibinfo {author} {\bibfnamefont {J.-W.}\
			\bibnamefont {Pan}},\ }\bibfield  {title} {\bibinfo {title} {Dynamic
			formation of quasicondensate and spontaneous vortices in a strongly
			interacting fermi gas},\ }\href
	{https://doi.org/10.1103/PhysRevResearch.3.043115} {\bibfield  {journal}
		{\bibinfo  {journal} {Phys. Rev. Res.}\ }\textbf {\bibinfo {volume} {3}},\
		\bibinfo {pages} {043115} (\bibinfo {year} {2021})}\BibitemShut {NoStop}%
	\bibitem [{\citenamefont {He}\ and\ \citenamefont {Chien}(2023)}]{HePLA23}%
	\BibitemOpen
	\bibfield  {author} {\bibinfo {author} {\bibfnamefont {Y.}~\bibnamefont
			{He}}\ and\ \bibinfo {author} {\bibfnamefont {C.-C.}\ \bibnamefont {Chien}},\
	}\bibfield  {title} {\bibinfo {title} {Particle and thermal transport through
			one dimensional topological systems via lindblad formalism},\ }\href
	{https://doi.org/https://doi.org/10.1016/j.physleta.2023.128826} {\bibfield
		{journal} {\bibinfo  {journal} {Physics Letters A}\ }\textbf {\bibinfo
			{volume} {473}},\ \bibinfo {pages} {128826} (\bibinfo {year}
		{2023})}\BibitemShut {NoStop}%
	\bibitem [{\citenamefont {Molignini}\ \emph {et~al.}(2018)\citenamefont
		{Molignini}, \citenamefont {Chen},\ and\ \citenamefont
		{Chitra}}]{PhysRevB.98.125129}%
	\BibitemOpen
	\bibfield  {author} {\bibinfo {author} {\bibfnamefont {P.}~\bibnamefont
			{Molignini}}, \bibinfo {author} {\bibfnamefont {W.}~\bibnamefont {Chen}},\
		and\ \bibinfo {author} {\bibfnamefont {R.}~\bibnamefont {Chitra}},\
	}\bibfield  {title} {\bibinfo {title} {Universal quantum criticality in
			static and floquet-majorana chains},\ }\href
	{https://doi.org/10.1103/PhysRevB.98.125129} {\bibfield  {journal} {\bibinfo
			{journal} {Phys. Rev. B}\ }\textbf {\bibinfo {volume} {98}},\ \bibinfo
		{pages} {125129} (\bibinfo {year} {2018})}\BibitemShut {NoStop}%
	\bibitem [{\citenamefont {Lieb}\ \emph {et~al.}(1961)\citenamefont {Lieb},
		\citenamefont {Schultz},\ and\ \citenamefont {Mattis}}]{Lieb61}%
	\BibitemOpen
	\bibfield  {author} {\bibinfo {author} {\bibfnamefont {E.}~\bibnamefont
			{Lieb}}, \bibinfo {author} {\bibfnamefont {T.}~\bibnamefont {Schultz}},\ and\
		\bibinfo {author} {\bibfnamefont {D.}~\bibnamefont {Mattis}},\ }\bibfield
	{title} {\bibinfo {title} {Two soluble models of an antiferromagnetic
			chain},\ }\href
	{https://doi.org/https://doi.org/10.1016/0003-4916(61)90115-4} {\bibfield
		{journal} {\bibinfo  {journal} {Annals of Physics}\ }\textbf {\bibinfo
			{volume} {16}},\ \bibinfo {pages} {407} (\bibinfo {year} {1961})}\BibitemShut
	{NoStop}%
	\bibitem [{\citenamefont {Deutscher}\ and\ \citenamefont
		{de~Gennes}(1969)}]{degennes-proximity}%
	\BibitemOpen
	\bibfield  {author} {\bibinfo {author} {\bibfnamefont {G.}~\bibnamefont
			{Deutscher}}\ and\ \bibinfo {author} {\bibfnamefont {P.}~\bibnamefont
			{de~Gennes}},\ }\bibfield  {title} {{\selectlanguage {english}\bibinfo
			{title} {Proximity effects}},\ }\href@noop {} {\bibfield  {journal} {\bibinfo
			{journal} {pp 1005-34 of Superconductivity. Vols. 1 and 2. Parks, R. D.
				(ed.). New York, Marcel Dekker, Inc., 1969}\ } (\bibinfo {year}
		{1969})}\BibitemShut {NoStop}%
	\bibitem [{\citenamefont {Falk}(1963)}]{Falk-PE}%
	\BibitemOpen
	\bibfield  {author} {\bibinfo {author} {\bibfnamefont {D.~S.}\ \bibnamefont
			{Falk}},\ }\bibfield  {title} {\bibinfo {title} {Superconductors with plane
			boundaries},\ }\href {https://doi.org/10.1103/PhysRev.132.1576} {\bibfield
		{journal} {\bibinfo  {journal} {Phys. Rev.}\ }\textbf {\bibinfo {volume}
			{132}},\ \bibinfo {pages} {1576} (\bibinfo {year} {1963})}\BibitemShut
	{NoStop}%
	\bibitem [{\citenamefont {Silvert}(1964)}]{PE-Silvert}%
	\BibitemOpen
	\bibfield  {author} {\bibinfo {author} {\bibfnamefont {W.}~\bibnamefont
			{Silvert}},\ }\bibfield  {title} {\bibinfo {title} {Spatial dependence of
			pair correlation functions in nonhomogeneous superconductors},\ }\href
	{https://doi.org/10.1103/RevModPhys.36.251} {\bibfield  {journal} {\bibinfo
			{journal} {Rev. Mod. Phys.}\ }\textbf {\bibinfo {volume} {36}},\ \bibinfo
		{pages} {251} (\bibinfo {year} {1964})}\BibitemShut {NoStop}%
	\bibitem [{\citenamefont {Deutsch}(2018)}]{Deutsch_2018}%
	\BibitemOpen
	\bibfield  {author} {\bibinfo {author} {\bibfnamefont {J.~M.}\ \bibnamefont
			{Deutsch}},\ }\bibfield  {title} {\bibinfo {title} {Eigenstate thermalization
			hypothesis},\ }\href {https://doi.org/10.1088/1361-6633/aac9f1} {\bibfield
		{journal} {\bibinfo  {journal} {Reports on Progress in Physics}\ }\textbf
		{\bibinfo {volume} {81}},\ \bibinfo {pages} {082001} (\bibinfo {year}
		{2018})}\BibitemShut {NoStop}%
	\bibitem [{\citenamefont {Altman}(2018)}]{Altman2018}%
	\BibitemOpen
	\bibfield  {author} {\bibinfo {author} {\bibfnamefont {E.}~\bibnamefont
			{Altman}},\ }\bibfield  {title} {\bibinfo {title} {Many-body localization and
			quantum thermalization},\ }\href {https://doi.org/10.1038/s41567-018-0305-7}
	{\bibfield  {journal} {\bibinfo  {journal} {Nature Physics}\ }\textbf
		{\bibinfo {volume} {14}},\ \bibinfo {pages} {979} (\bibinfo {year}
		{2018})}\BibitemShut {NoStop}%
	\bibitem [{\citenamefont {Abanin}\ \emph {et~al.}(2019)\citenamefont {Abanin},
		\citenamefont {Altman}, \citenamefont {Bloch},\ and\ \citenamefont
		{Serbyn}}]{RevModPhys.91.021001}%
	\BibitemOpen
	\bibfield  {author} {\bibinfo {author} {\bibfnamefont {D.~A.}\ \bibnamefont
			{Abanin}}, \bibinfo {author} {\bibfnamefont {E.}~\bibnamefont {Altman}},
		\bibinfo {author} {\bibfnamefont {I.}~\bibnamefont {Bloch}},\ and\ \bibinfo
		{author} {\bibfnamefont {M.}~\bibnamefont {Serbyn}},\ }\bibfield  {title}
	{\bibinfo {title} {Colloquium: Many-body localization, thermalization, and
			entanglement},\ }\href {https://doi.org/10.1103/RevModPhys.91.021001}
	{\bibfield  {journal} {\bibinfo  {journal} {Rev. Mod. Phys.}\ }\textbf
		{\bibinfo {volume} {91}},\ \bibinfo {pages} {021001} (\bibinfo {year}
		{2019})}\BibitemShut {NoStop}%
	\bibitem [{\citenamefont {Dvir}\ \emph {et~al.}(2023)\citenamefont {Dvir},
		\citenamefont {Wang}, \citenamefont {van Loo}, \citenamefont {Liu},
		\citenamefont {Mazur}, \citenamefont {Bordin}, \citenamefont {ten Haaf},
		\citenamefont {Wang}, \citenamefont {van Driel}, \citenamefont {Zatelli},
		\citenamefont {Li}, \citenamefont {Malinowski}, \citenamefont {Gazibegovic},
		\citenamefont {Badawy}, \citenamefont {Bakkers}, \citenamefont {Wimmer},\
		and\ \citenamefont {Kouwenhoven}}]{Dvir23}%
	\BibitemOpen
	\bibfield  {author} {\bibinfo {author} {\bibfnamefont {T.}~\bibnamefont
			{Dvir}}, \bibinfo {author} {\bibfnamefont {G.}~\bibnamefont {Wang}}, \bibinfo
		{author} {\bibfnamefont {N.}~\bibnamefont {van Loo}}, \bibinfo {author}
		{\bibfnamefont {C.-X.}\ \bibnamefont {Liu}}, \bibinfo {author} {\bibfnamefont
			{G.~P.}\ \bibnamefont {Mazur}}, \bibinfo {author} {\bibfnamefont
			{A.}~\bibnamefont {Bordin}}, \bibinfo {author} {\bibfnamefont {S.~L.~D.}\
			\bibnamefont {ten Haaf}}, \bibinfo {author} {\bibfnamefont {J.-Y.}\
			\bibnamefont {Wang}}, \bibinfo {author} {\bibfnamefont {D.}~\bibnamefont {van
				Driel}}, \bibinfo {author} {\bibfnamefont {F.}~\bibnamefont {Zatelli}},
		\bibinfo {author} {\bibfnamefont {X.}~\bibnamefont {Li}}, \bibinfo {author}
		{\bibfnamefont {F.~K.}\ \bibnamefont {Malinowski}}, \bibinfo {author}
		{\bibfnamefont {S.}~\bibnamefont {Gazibegovic}}, \bibinfo {author}
		{\bibfnamefont {G.}~\bibnamefont {Badawy}}, \bibinfo {author} {\bibfnamefont
			{E.~P. A.~M.}\ \bibnamefont {Bakkers}}, \bibinfo {author} {\bibfnamefont
			{M.}~\bibnamefont {Wimmer}},\ and\ \bibinfo {author} {\bibfnamefont {L.~P.}\
			\bibnamefont {Kouwenhoven}},\ }\bibfield  {title} {\bibinfo {title}
		{Realization of a minimal kitaev chain in coupled quantum dots},\ }\href
	{https://doi.org/10.1038/s41586-022-05585-1} {\bibfield  {journal} {\bibinfo
			{journal} {Nature}\ }\textbf {\bibinfo {volume} {614}},\ \bibinfo {pages}
		{445} (\bibinfo {year} {2023})}\BibitemShut {NoStop}%
	\bibitem [{\citenamefont {Pan}\ and\ \citenamefont
		{Das~Sarma}(2023)}]{PhysRevB.107.035440}%
	\BibitemOpen
	\bibfield  {author} {\bibinfo {author} {\bibfnamefont {H.}~\bibnamefont
			{Pan}}\ and\ \bibinfo {author} {\bibfnamefont {S.}~\bibnamefont
			{Das~Sarma}},\ }\bibfield  {title} {\bibinfo {title} {Majorana nanowires,
			kitaev chains, and spin models},\ }\href
	{https://doi.org/10.1103/PhysRevB.107.035440} {\bibfield  {journal} {\bibinfo
			{journal} {Phys. Rev. B}\ }\textbf {\bibinfo {volume} {107}},\ \bibinfo
		{pages} {035440} (\bibinfo {year} {2023})}\BibitemShut {NoStop}%
	\bibitem [{\citenamefont {Oppen}\ \emph {et~al.}(2017)\citenamefont {Oppen},
		\citenamefont {Peng},\ and\ \citenamefont {Pientka}}]{Oppen17}%
	\BibitemOpen
	\bibfield  {author} {\bibinfo {author} {\bibfnamefont {F.~v.}\ \bibnamefont
			{Oppen}}, \bibinfo {author} {\bibfnamefont {Y.}~\bibnamefont {Peng}},\ and\
		\bibinfo {author} {\bibfnamefont {F.}~\bibnamefont {Pientka}},\ }\bibfield
	{title} {\bibinfo {title} {{Topological superconducting phases in one
				dimension}},\ }in\ \href
	{https://doi.org/10.1093/acprof:oso/9780198785781.003.0009} {\emph {\bibinfo
			{booktitle} {{Topological Aspects of Condensed Matter Physics: Lecture Notes
					of the Les Houches Summer School: Volume 103, August 2014}}}}\ (\bibinfo
	{publisher} {Oxford University Press},\ \bibinfo {year} {2017})\ \Eprint
	{https://arxiv.org/abs/https://academic.oup.com/book/0/chapter/203983732/chapter-pdf/45122673/acprof-9780198785781-chapter-9.pdf}
	{https://academic.oup.com/book/0/chapter/203983732/chapter-pdf/45122673/acprof-9780198785781-chapter-9.pdf}
	\BibitemShut {NoStop}%
	\bibitem [{\citenamefont {Stenger}\ \emph {et~al.}(2021)\citenamefont
		{Stenger}, \citenamefont {Bronn}, \citenamefont {Egger},\ and\ \citenamefont
		{Pekker}}]{Stenger21}%
	\BibitemOpen
	\bibfield  {author} {\bibinfo {author} {\bibfnamefont {J.~P.~T.}\
			\bibnamefont {Stenger}}, \bibinfo {author} {\bibfnamefont {N.~T.}\
			\bibnamefont {Bronn}}, \bibinfo {author} {\bibfnamefont {D.~J.}\ \bibnamefont
			{Egger}},\ and\ \bibinfo {author} {\bibfnamefont {D.}~\bibnamefont
			{Pekker}},\ }\bibfield  {title} {\bibinfo {title} {Simulating the dynamics of
			braiding of majorana zero modes using an ibm quantum computer},\ }\href@noop
	{} {\bibfield  {journal} {\bibinfo  {journal} {Phys. Rev. Research}\ }\textbf
		{\bibinfo {volume} {3}},\ \bibinfo {pages} {033171} (\bibinfo {year}
		{2021})}\BibitemShut {NoStop}%
	\bibitem [{\citenamefont {Huang}\ \emph {et~al.}(2021)\citenamefont {Huang},
		\citenamefont {Naro\ifmmode~\dot{z}\else \.{z}\fi{}niak}, \citenamefont
		{Liang}, \citenamefont {Zhao}, \citenamefont {Castellano}, \citenamefont
		{Gong}, \citenamefont {Wu}, \citenamefont {Wang}, \citenamefont {Lin},
		\citenamefont {Xu}, \citenamefont {Deng}, \citenamefont {Rong}, \citenamefont
		{Dowling}, \citenamefont {Peng}, \citenamefont {Byrnes}, \citenamefont
		{Zhu},\ and\ \citenamefont {Pan}}]{Huang21}%
	\BibitemOpen
	\bibfield  {author} {\bibinfo {author} {\bibfnamefont {H.-L.}\ \bibnamefont
			{Huang}}, \bibinfo {author} {\bibfnamefont {M.}~\bibnamefont
			{Naro\ifmmode~\dot{z}\else \.{z}\fi{}niak}}, \bibinfo {author} {\bibfnamefont
			{F.}~\bibnamefont {Liang}}, \bibinfo {author} {\bibfnamefont
			{Y.}~\bibnamefont {Zhao}}, \bibinfo {author} {\bibfnamefont {A.~D.}\
			\bibnamefont {Castellano}}, \bibinfo {author} {\bibfnamefont
			{M.}~\bibnamefont {Gong}}, \bibinfo {author} {\bibfnamefont {Y.}~\bibnamefont
			{Wu}}, \bibinfo {author} {\bibfnamefont {S.}~\bibnamefont {Wang}}, \bibinfo
		{author} {\bibfnamefont {J.}~\bibnamefont {Lin}}, \bibinfo {author}
		{\bibfnamefont {Y.}~\bibnamefont {Xu}}, \bibinfo {author} {\bibfnamefont
			{H.}~\bibnamefont {Deng}}, \bibinfo {author} {\bibfnamefont {H.}~\bibnamefont
			{Rong}}, \bibinfo {author} {\bibfnamefont {J.~P.}\ \bibnamefont {Dowling}},
		\bibinfo {author} {\bibfnamefont {C.-Z.}\ \bibnamefont {Peng}}, \bibinfo
		{author} {\bibfnamefont {T.}~\bibnamefont {Byrnes}}, \bibinfo {author}
		{\bibfnamefont {X.}~\bibnamefont {Zhu}},\ and\ \bibinfo {author}
		{\bibfnamefont {J.-W.}\ \bibnamefont {Pan}},\ }\bibfield  {title} {\bibinfo
		{title} {Emulating quantum teleportation of a majorana zero mode qubit},\
	}\href {https://doi.org/10.1103/PhysRevLett.126.090502} {\bibfield  {journal}
		{\bibinfo  {journal} {Phys. Rev. Lett.}\ }\textbf {\bibinfo {volume} {126}},\
		\bibinfo {pages} {090502} (\bibinfo {year} {2021})}\BibitemShut {NoStop}%
	\bibitem [{\citenamefont {Mi}\ \emph {et~al.}(2022)\citenamefont {Mi},
		\citenamefont {Sonner}, \citenamefont {Niu}, \citenamefont {Lee},
		\citenamefont {Foxen}, \citenamefont {Acharya}, \citenamefont {Aleiner},
		\citenamefont {Andersen}, \citenamefont {Arute}, \citenamefont {Arya},
		\citenamefont {Asfaw}, \citenamefont {Atalaya}, \citenamefont {Bardin},
		\citenamefont {Basso}, \citenamefont {Bengtsson}, \citenamefont {Bortoli},
		\citenamefont {Bourassa}, \citenamefont {Brill}, \citenamefont {Broughton},
		\citenamefont {Buckley}, \citenamefont {Buell}, \citenamefont {Burkett},
		\citenamefont {Bushnell}, \citenamefont {Chen}, \citenamefont {Chiaro},
		\citenamefont {Collins}, \citenamefont {Conner}, \citenamefont {Courtney},
		\citenamefont {Crook}, \citenamefont {Debroy}, \citenamefont {Demura},
		\citenamefont {Dunsworth}, \citenamefont {Eppens}, \citenamefont {Erickson},
		\citenamefont {Faoro}, \citenamefont {Farhi}, \citenamefont {Fatemi},
		\citenamefont {Flores}, \citenamefont {Forati}, \citenamefont {Fowler},
		\citenamefont {Giang}, \citenamefont {Gidney}, \citenamefont {Gilboa},
		\citenamefont {Giustina}, \citenamefont {Dau}, \citenamefont {Gross},
		\citenamefont {Habegger}, \citenamefont {Harrigan}, \citenamefont {Hoffmann},
		\citenamefont {Hong}, \citenamefont {Huang}, \citenamefont {Huff},
		\citenamefont {Huggins}, \citenamefont {Ioffe}, \citenamefont {Isakov},
		\citenamefont {Iveland}, \citenamefont {Jeffrey}, \citenamefont {Jiang},
		\citenamefont {Jones}, \citenamefont {Kafri}, \citenamefont {Kechedzhi},
		\citenamefont {Khattar}, \citenamefont {Kim}, \citenamefont {Kitaev},
		\citenamefont {Klimov}, \citenamefont {Klots}, \citenamefont {Korotkov},
		\citenamefont {Kostritsa}, \citenamefont {Kreikebaum}, \citenamefont
		{Landhuis}, \citenamefont {Laptev}, \citenamefont {Lau}, \citenamefont {Lee},
		\citenamefont {Laws}, \citenamefont {Liu}, \citenamefont {Locharla},
		\citenamefont {Martin}, \citenamefont {McClean}, \citenamefont {McEwen},
		\citenamefont {Costa}, \citenamefont {Miao}, \citenamefont {Mohseni},
		\citenamefont {Montazeri}, \citenamefont {Morvan}, \citenamefont {Mount},
		\citenamefont {Mruczkiewicz}, \citenamefont {Naaman}, \citenamefont {Neeley},
		\citenamefont {Neill}, \citenamefont {Newman}, \citenamefont {O’Brien},
		\citenamefont {Opremcak}, \citenamefont {Petukhov}, \citenamefont {Potter},
		\citenamefont {Quintana}, \citenamefont {Rubin}, \citenamefont {Saei},
		\citenamefont {Sank}, \citenamefont {Sankaragomathi}, \citenamefont
		{Satzinger}, \citenamefont {Schuster}, \citenamefont {Shearn}, \citenamefont
		{Shvarts}, \citenamefont {Strain}, \citenamefont {Su}, \citenamefont
		{Szalay}, \citenamefont {Vidal}, \citenamefont {Villalonga}, \citenamefont
		{Vollgraff-Heidweiller}, \citenamefont {White}, \citenamefont {Yao},
		\citenamefont {Yeh}, \citenamefont {Yoo}, \citenamefont {Zalcman},
		\citenamefont {Zhang}, \citenamefont {Zhu}, \citenamefont {Neven},
		\citenamefont {Bacon}, \citenamefont {Hilton}, \citenamefont {Lucero},
		\citenamefont {Babbush}, \citenamefont {Boixo}, \citenamefont {Megrant},
		\citenamefont {Chen}, \citenamefont {Kelly}, \citenamefont {Smelyanskiy},
		\citenamefont {Abanin},\ and\ \citenamefont {Roushan}}]{Mi22}%
	\BibitemOpen
	\bibfield  {author} {\bibinfo {author} {\bibfnamefont {X.}~\bibnamefont
			{Mi}}, \bibinfo {author} {\bibfnamefont {M.}~\bibnamefont {Sonner}}, \bibinfo
		{author} {\bibfnamefont {M.~Y.}\ \bibnamefont {Niu}}, \bibinfo {author}
		{\bibfnamefont {K.~W.}\ \bibnamefont {Lee}}, \bibinfo {author} {\bibfnamefont
			{B.}~\bibnamefont {Foxen}}, \bibinfo {author} {\bibfnamefont
			{R.}~\bibnamefont {Acharya}}, \bibinfo {author} {\bibfnamefont
			{I.}~\bibnamefont {Aleiner}}, \bibinfo {author} {\bibfnamefont {T.~I.}\
			\bibnamefont {Andersen}}, \bibinfo {author} {\bibfnamefont {F.}~\bibnamefont
			{Arute}}, \bibinfo {author} {\bibfnamefont {K.}~\bibnamefont {Arya}},
		\bibinfo {author} {\bibfnamefont {A.}~\bibnamefont {Asfaw}}, \bibinfo
		{author} {\bibfnamefont {J.}~\bibnamefont {Atalaya}}, \bibinfo {author}
		{\bibfnamefont {J.~C.}\ \bibnamefont {Bardin}}, \bibinfo {author}
		{\bibfnamefont {J.}~\bibnamefont {Basso}}, \bibinfo {author} {\bibfnamefont
			{A.}~\bibnamefont {Bengtsson}}, \bibinfo {author} {\bibfnamefont
			{G.}~\bibnamefont {Bortoli}}, \bibinfo {author} {\bibfnamefont
			{A.}~\bibnamefont {Bourassa}}, \bibinfo {author} {\bibfnamefont
			{L.}~\bibnamefont {Brill}}, \bibinfo {author} {\bibfnamefont
			{M.}~\bibnamefont {Broughton}}, \bibinfo {author} {\bibfnamefont {B.~B.}\
			\bibnamefont {Buckley}}, \bibinfo {author} {\bibfnamefont {D.~A.}\
			\bibnamefont {Buell}}, \bibinfo {author} {\bibfnamefont {B.}~\bibnamefont
			{Burkett}}, \bibinfo {author} {\bibfnamefont {N.}~\bibnamefont {Bushnell}},
		\bibinfo {author} {\bibfnamefont {Z.}~\bibnamefont {Chen}}, \bibinfo {author}
		{\bibfnamefont {B.}~\bibnamefont {Chiaro}}, \bibinfo {author} {\bibfnamefont
			{R.}~\bibnamefont {Collins}}, \bibinfo {author} {\bibfnamefont
			{P.}~\bibnamefont {Conner}}, \bibinfo {author} {\bibfnamefont
			{W.}~\bibnamefont {Courtney}}, \bibinfo {author} {\bibfnamefont {A.~L.}\
			\bibnamefont {Crook}}, \bibinfo {author} {\bibfnamefont {D.~M.}\ \bibnamefont
			{Debroy}}, \bibinfo {author} {\bibfnamefont {S.}~\bibnamefont {Demura}},
		\bibinfo {author} {\bibfnamefont {A.}~\bibnamefont {Dunsworth}}, \bibinfo
		{author} {\bibfnamefont {D.}~\bibnamefont {Eppens}}, \bibinfo {author}
		{\bibfnamefont {C.}~\bibnamefont {Erickson}}, \bibinfo {author}
		{\bibfnamefont {L.}~\bibnamefont {Faoro}}, \bibinfo {author} {\bibfnamefont
			{E.}~\bibnamefont {Farhi}}, \bibinfo {author} {\bibfnamefont
			{R.}~\bibnamefont {Fatemi}}, \bibinfo {author} {\bibfnamefont
			{L.}~\bibnamefont {Flores}}, \bibinfo {author} {\bibfnamefont
			{E.}~\bibnamefont {Forati}}, \bibinfo {author} {\bibfnamefont {A.~G.}\
			\bibnamefont {Fowler}}, \bibinfo {author} {\bibfnamefont {W.}~\bibnamefont
			{Giang}}, \bibinfo {author} {\bibfnamefont {C.}~\bibnamefont {Gidney}},
		\bibinfo {author} {\bibfnamefont {D.}~\bibnamefont {Gilboa}}, \bibinfo
		{author} {\bibfnamefont {M.}~\bibnamefont {Giustina}}, \bibinfo {author}
		{\bibfnamefont {A.~G.}\ \bibnamefont {Dau}}, \bibinfo {author} {\bibfnamefont
			{J.~A.}\ \bibnamefont {Gross}}, \bibinfo {author} {\bibfnamefont
			{S.}~\bibnamefont {Habegger}}, \bibinfo {author} {\bibfnamefont {M.~P.}\
			\bibnamefont {Harrigan}}, \bibinfo {author} {\bibfnamefont {M.}~\bibnamefont
			{Hoffmann}}, \bibinfo {author} {\bibfnamefont {S.}~\bibnamefont {Hong}},
		\bibinfo {author} {\bibfnamefont {T.}~\bibnamefont {Huang}}, \bibinfo
		{author} {\bibfnamefont {A.}~\bibnamefont {Huff}}, \bibinfo {author}
		{\bibfnamefont {W.~J.}\ \bibnamefont {Huggins}}, \bibinfo {author}
		{\bibfnamefont {L.~B.}\ \bibnamefont {Ioffe}}, \bibinfo {author}
		{\bibfnamefont {S.~V.}\ \bibnamefont {Isakov}}, \bibinfo {author}
		{\bibfnamefont {J.}~\bibnamefont {Iveland}}, \bibinfo {author} {\bibfnamefont
			{E.}~\bibnamefont {Jeffrey}}, \bibinfo {author} {\bibfnamefont
			{Z.}~\bibnamefont {Jiang}}, \bibinfo {author} {\bibfnamefont
			{C.}~\bibnamefont {Jones}}, \bibinfo {author} {\bibfnamefont
			{D.}~\bibnamefont {Kafri}}, \bibinfo {author} {\bibfnamefont
			{K.}~\bibnamefont {Kechedzhi}}, \bibinfo {author} {\bibfnamefont
			{T.}~\bibnamefont {Khattar}}, \bibinfo {author} {\bibfnamefont
			{S.}~\bibnamefont {Kim}}, \bibinfo {author} {\bibfnamefont {A.~Y.}\
			\bibnamefont {Kitaev}}, \bibinfo {author} {\bibfnamefont {P.~V.}\
			\bibnamefont {Klimov}}, \bibinfo {author} {\bibfnamefont {A.~R.}\
			\bibnamefont {Klots}}, \bibinfo {author} {\bibfnamefont {A.~N.}\ \bibnamefont
			{Korotkov}}, \bibinfo {author} {\bibfnamefont {F.}~\bibnamefont {Kostritsa}},
		\bibinfo {author} {\bibfnamefont {J.~M.}\ \bibnamefont {Kreikebaum}},
		\bibinfo {author} {\bibfnamefont {D.}~\bibnamefont {Landhuis}}, \bibinfo
		{author} {\bibfnamefont {P.}~\bibnamefont {Laptev}}, \bibinfo {author}
		{\bibfnamefont {K.-M.}\ \bibnamefont {Lau}}, \bibinfo {author} {\bibfnamefont
			{J.}~\bibnamefont {Lee}}, \bibinfo {author} {\bibfnamefont {L.}~\bibnamefont
			{Laws}}, \bibinfo {author} {\bibfnamefont {W.}~\bibnamefont {Liu}}, \bibinfo
		{author} {\bibfnamefont {A.}~\bibnamefont {Locharla}}, \bibinfo {author}
		{\bibfnamefont {O.}~\bibnamefont {Martin}}, \bibinfo {author} {\bibfnamefont
			{J.~R.}\ \bibnamefont {McClean}}, \bibinfo {author} {\bibfnamefont
			{M.}~\bibnamefont {McEwen}}, \bibinfo {author} {\bibfnamefont {B.~M.}\
			\bibnamefont {Costa}}, \bibinfo {author} {\bibfnamefont {K.~C.}\ \bibnamefont
			{Miao}}, \bibinfo {author} {\bibfnamefont {M.}~\bibnamefont {Mohseni}},
		\bibinfo {author} {\bibfnamefont {S.}~\bibnamefont {Montazeri}}, \bibinfo
		{author} {\bibfnamefont {A.}~\bibnamefont {Morvan}}, \bibinfo {author}
		{\bibfnamefont {E.}~\bibnamefont {Mount}}, \bibinfo {author} {\bibfnamefont
			{W.}~\bibnamefont {Mruczkiewicz}}, \bibinfo {author} {\bibfnamefont
			{O.}~\bibnamefont {Naaman}}, \bibinfo {author} {\bibfnamefont
			{M.}~\bibnamefont {Neeley}}, \bibinfo {author} {\bibfnamefont
			{C.}~\bibnamefont {Neill}}, \bibinfo {author} {\bibfnamefont
			{M.}~\bibnamefont {Newman}}, \bibinfo {author} {\bibfnamefont {T.~E.}\
			\bibnamefont {O’Brien}}, \bibinfo {author} {\bibfnamefont {A.}~\bibnamefont
			{Opremcak}}, \bibinfo {author} {\bibfnamefont {A.}~\bibnamefont {Petukhov}},
		\bibinfo {author} {\bibfnamefont {R.}~\bibnamefont {Potter}}, \bibinfo
		{author} {\bibfnamefont {C.}~\bibnamefont {Quintana}}, \bibinfo {author}
		{\bibfnamefont {N.~C.}\ \bibnamefont {Rubin}}, \bibinfo {author}
		{\bibfnamefont {N.}~\bibnamefont {Saei}}, \bibinfo {author} {\bibfnamefont
			{D.}~\bibnamefont {Sank}}, \bibinfo {author} {\bibfnamefont {K.}~\bibnamefont
			{Sankaragomathi}}, \bibinfo {author} {\bibfnamefont {K.~J.}\ \bibnamefont
			{Satzinger}}, \bibinfo {author} {\bibfnamefont {C.}~\bibnamefont {Schuster}},
		\bibinfo {author} {\bibfnamefont {M.~J.}\ \bibnamefont {Shearn}}, \bibinfo
		{author} {\bibfnamefont {V.}~\bibnamefont {Shvarts}}, \bibinfo {author}
		{\bibfnamefont {D.}~\bibnamefont {Strain}}, \bibinfo {author} {\bibfnamefont
			{Y.}~\bibnamefont {Su}}, \bibinfo {author} {\bibfnamefont {M.}~\bibnamefont
			{Szalay}}, \bibinfo {author} {\bibfnamefont {G.}~\bibnamefont {Vidal}},
		\bibinfo {author} {\bibfnamefont {B.}~\bibnamefont {Villalonga}}, \bibinfo
		{author} {\bibfnamefont {C.}~\bibnamefont {Vollgraff-Heidweiller}}, \bibinfo
		{author} {\bibfnamefont {T.}~\bibnamefont {White}}, \bibinfo {author}
		{\bibfnamefont {Z.}~\bibnamefont {Yao}}, \bibinfo {author} {\bibfnamefont
			{P.}~\bibnamefont {Yeh}}, \bibinfo {author} {\bibfnamefont {J.}~\bibnamefont
			{Yoo}}, \bibinfo {author} {\bibfnamefont {A.}~\bibnamefont {Zalcman}},
		\bibinfo {author} {\bibfnamefont {Y.}~\bibnamefont {Zhang}}, \bibinfo
		{author} {\bibfnamefont {N.}~\bibnamefont {Zhu}}, \bibinfo {author}
		{\bibfnamefont {H.}~\bibnamefont {Neven}}, \bibinfo {author} {\bibfnamefont
			{D.}~\bibnamefont {Bacon}}, \bibinfo {author} {\bibfnamefont
			{J.}~\bibnamefont {Hilton}}, \bibinfo {author} {\bibfnamefont
			{E.}~\bibnamefont {Lucero}}, \bibinfo {author} {\bibfnamefont
			{R.}~\bibnamefont {Babbush}}, \bibinfo {author} {\bibfnamefont
			{S.}~\bibnamefont {Boixo}}, \bibinfo {author} {\bibfnamefont
			{A.}~\bibnamefont {Megrant}}, \bibinfo {author} {\bibfnamefont
			{Y.}~\bibnamefont {Chen}}, \bibinfo {author} {\bibfnamefont {J.}~\bibnamefont
			{Kelly}}, \bibinfo {author} {\bibfnamefont {V.}~\bibnamefont {Smelyanskiy}},
		\bibinfo {author} {\bibfnamefont {D.~A.}\ \bibnamefont {Abanin}},\ and\
		\bibinfo {author} {\bibfnamefont {P.}~\bibnamefont {Roushan}},\ }\bibfield
	{title} {\bibinfo {title} {Noise-resilient edge modes on a chain of
			superconducting qubits},\ }\href@noop {} {\bibfield  {journal} {\bibinfo
			{journal} {Science}\ }\textbf {\bibinfo {volume} {378}},\ \bibinfo {pages}
		{785} (\bibinfo {year} {2022})}\BibitemShut {NoStop}%
	\bibitem [{\citenamefont {Ran{\v{c}}i{\'{c}}}(2022)}]{Rancic2022}%
	\BibitemOpen
	\bibfield  {author} {\bibinfo {author} {\bibfnamefont {M.~J.}\ \bibnamefont
			{Ran{\v{c}}i{\'{c}}}},\ }\bibfield  {title} {\bibinfo {title} {Exactly
			solving the kitaev chain and generating majorana-zero-modes out of noisy
			qubits},\ }\href {https://doi.org/10.1038/s41598-022-24341-z} {\bibfield
		{journal} {\bibinfo  {journal} {Sci. Rep.}\ }\textbf {\bibinfo {volume}
			{12}},\ \bibinfo {pages} {19882} (\bibinfo {year} {2022})}\BibitemShut
	{NoStop}%
	\bibitem [{\citenamefont {Iizuka}\ \emph {et~al.}(2023)\citenamefont {Iizuka},
		\citenamefont {Yuan}, \citenamefont {Mita}, \citenamefont {Higo},
		\citenamefont {Yasunaga},\ and\ \citenamefont {Ezawa}}]{Iizuka2023}%
	\BibitemOpen
	\bibfield  {author} {\bibinfo {author} {\bibfnamefont {T.}~\bibnamefont
			{Iizuka}}, \bibinfo {author} {\bibfnamefont {H.}~\bibnamefont {Yuan}},
		\bibinfo {author} {\bibfnamefont {Y.}~\bibnamefont {Mita}}, \bibinfo {author}
		{\bibfnamefont {A.}~\bibnamefont {Higo}}, \bibinfo {author} {\bibfnamefont
			{S.}~\bibnamefont {Yasunaga}},\ and\ \bibinfo {author} {\bibfnamefont
			{M.}~\bibnamefont {Ezawa}},\ }\bibfield  {title} {\bibinfo {title}
		{Experimental demonstration of position-controllable topological interface
			states in high-frequency kitaev topological integrated circuits},\ }\href
	{https://doi.org/10.1038/s42005-023-01404-9} {\bibfield  {journal} {\bibinfo
			{journal} {Commun. Phys.}\ }\textbf {\bibinfo {volume} {6}},\ \bibinfo
		{pages} {279} (\bibinfo {year} {2023})}\BibitemShut {NoStop}%
	\bibitem [{\citenamefont {Buzdin}(2005)}]{SC-FM-rev}%
	\BibitemOpen
	\bibfield  {author} {\bibinfo {author} {\bibfnamefont {A.~I.}\ \bibnamefont
			{Buzdin}},\ }\bibfield  {title} {\bibinfo {title} {Proximity effects in
			superconductor-ferromagnet heterostructures},\ }\href
	{https://doi.org/10.1103/RevModPhys.77.935} {\bibfield  {journal} {\bibinfo
			{journal} {Rev. Mod. Phys.}\ }\textbf {\bibinfo {volume} {77}},\ \bibinfo
		{pages} {935} (\bibinfo {year} {2005})}\BibitemShut {NoStop}%
\end{thebibliography}
%

\end{document}